\documentclass[11pt,a4paper]{article}
\usepackage[margin=1in]{geometry}
\usepackage{amsmath,amssymb,amsthm,mathtools}
\usepackage{enumitem}
\usepackage{xcolor}
\usepackage{graphicx}
\definecolor{linkblue}{RGB}{0,70,140}
\usepackage{cite}
\usepackage[colorlinks=true,linkcolor=black,citecolor=black,urlcolor=black]{hyperref}
\usepackage{aliascnt}
\usepackage[nameinlink,noabbrev]{cleveref}
\numberwithin{equation}{section}
\usepackage{tabularx}
\usepackage{booktabs}
\theoremstyle{plain}
\newtheorem{theorem}{Theorem}[section]
\newaliascnt{proposition}{theorem}
\newtheorem{proposition}[proposition]{Proposition}
\aliascntresetthe{proposition}
\newaliascnt{lemma}{theorem}
\newtheorem{lemma}[lemma]{Lemma}
\aliascntresetthe{lemma}
\newaliascnt{corollary}{theorem}
\newtheorem{corollary}[corollary]{Corollary}
\aliascntresetthe{corollary}
\theoremstyle{definition}
\newaliascnt{definition}{theorem}
\newtheorem{definition}[definition]{Definition}
\aliascntresetthe{definition}
\theoremstyle{remark}
\newaliascnt{remark}{theorem}
\newtheorem{remark}[remark]{Remark}
\aliascntresetthe{remark}
\newaliascnt{example}{theorem}
\newtheorem{example}[example]{Example}
\aliascntresetthe{example}

\Crefname{theorem}{Theorem}{Theorems}
\Crefname{proposition}{Proposition}{Propositions}
\Crefname{lemma}{Lemma}{Lemmas}
\Crefname{corollary}{Corollary}{Corollaries}
\Crefname{definition}{Definition}{Definitions}
\Crefname{remark}{Remark}{Remarks}
\Crefname{example}{Example}{Examples}
\crefname{theorem}{Theorem}{Theorems}
\crefname{proposition}{Proposition}{Propositions}
\crefname{lemma}{Lemma}{Lemmas}
\crefname{corollary}{Corollary}{Corollaries}
\crefname{definition}{Definition}{Definitions}
\crefname{remark}{Remark}{Remarks}
\crefname{example}{Example}{Examples}
\crefname{section}{Section}{Sections}
\Crefname{section}{Section}{Sections}
\crefname{subsection}{Section}{Sections}
\Crefname{subsection}{Section}{Sections}
\crefname{subsubsection}{Section}{Sections}
\Crefname{subsubsection}{Section}{Sections}
\crefname{figure}{Figure}{Figures}
\Crefname{figure}{Figure}{Figures}
\crefname{table}{Table}{Tables}
\Crefname{table}{Table}{Tables}
\crefname{appendix}{Appendix}{Appendices}
\Crefname{appendix}{Appendix}{Appendices}

\newcommand{\FQ}{\mathbb F_Q}
\newcommand{\Ftwo}{\mathbb F_2}
\newcommand{\Prob}{\mathrm{Pr}}
\newcommand{\E}{\mathbb E}
\newcommand{\eps}{\varepsilon}
\newcommand{\Cin}{C_{\mathrm{in}}}
\newcommand{\Cout}{C_{\mathrm{out}}}
\newcommand{\Coutm}{C_{\mathrm{out},m}}
\newcommand{\wt}{\operatorname{wt}}
\newcommand{\supp}{\operatorname{supp}}
\newcommand{\diag}{\operatorname{diag}}
\newcommand{\1}{\mathbf 1}
\newcommand{\Tr}{\operatorname{Tr}}

\title{Concatenated Codes with One Inner Code: Low-Rate Gilbert-Varshamov Asymptotics from Algebraic-Geometric Codes}

\author{Yota Maeda\thanks{Email: \texttt{yota.maeda@mosk.tytlabs.co.jp}.}
\qquad
Ruho Kondo\thanks{Email: \texttt{r-kondo@mosk.tytlabs.co.jp}.}\\[0.5ex]
Toyota Central R\&D Labs., Inc.}
\date{}

\begin{document}
\maketitle

\begin{abstract}
Constructing codes that attain the Gilbert–Varshamov (GV) bound while admitting efficient encoding and decoding is a fundamental problem in coding theory. We study concatenated codes that reuse a single binary inner code across all outer coordinates. To exploit existing encoding and decoding algorithms, we restrict the outer codes to random subcodes of scaled copies of prescribed algebraic codes and investigate their minimum distance. Our first contribution is a general upper bound on the minimum-distance failure probability that separates two terms controlled by the inner code's bias moments and the dual weight distribution of the prescribed code, together with sufficient conditions for exponential decay. Our second contribution is a concrete construction satisfying these conditions using algebraic-geometric codes from the Garcia–Stichtenoth tower. For every sufficiently small $\varepsilon>0$, we construct a family of random binary codes $C_m$ whose rates approach $\varepsilon^2$ and satisfy
\[
\Pr\left[\delta(C_m)\geq\frac12-\sqrt{\frac{\ln 2}{2}}\left(\varepsilon+\frac{\varepsilon^3}{2}\right)\right]\geq 1-2^{-c\varepsilon^4N_m},
\]
for all sufficiently large $m$, where $N_m$ is the block length and $c>0$ is independent of $\varepsilon$ and $m$. Thus, the distance guarantee matches the leading coefficient of the GV bound at low rates. For each fixed $\varepsilon$, the construction uses a single inner encoding map throughout the family and admits polynomial-time encoding and unique decoding within a guaranteed radius. 
This answers the question of Doron, Mosheiff, and Wootters for subcodes of
scaled algebraic geometry codes at low rates.
\end{abstract}

\section{Introduction}
\label{sec:introduction}
\subsection{The GV bound for concatenated codes}
\label{subsec:gv-background}
The Gilbert--Varshamov (GV) bound gives a classical lower bound
on the best relative minimum distance attainable by binary codes at a given
rate \cite{Gilbert1952,Varshamov1957}.
Its distance function is $\delta_{\mathrm{GV}}(R)=h_2^{-1}(1-R)$, where $h_2(x):=-x\log_2x-(1-x)\log_2(1-x)$ is the binary entropy function (with $0\log_20:=0$), and $h_2^{-1}:[0,1]\to[0,1/2]$ is the inverse of the restriction $h_2|_{[0,1/2]}$.
A central problem in coding theory is to approach this bound with codes that admit efficient encoding and decoding.

Code concatenation is a classical approach to this problem.
In a binary concatenated code \cite{Forney1966}, each symbol of an
outer code over a larger finite field is encoded as a word of a binary
inner code.
The classical analysis based on the product of the inner and outer
minimum distances gives the Zyablov bound \cite{Zyablov1971}. Thommesen
showed that a Reed--Solomon outer code concatenated with independently chosen
random binary inner maps, one per outer coordinate, can attain the GV bound
\cite{Thommesen1983}.
Against this background, Doron, Mosheiff, and Wootters \cite{DoronMosheiffWootters2024} expressed high moments of the bias of concatenated codewords
in terms of the dual of the outer code. When both constituent codes are random
linear codes of rate $\eps$ and the outer alphabet is sufficiently large,
they obtain relative minimum distance at least $1/2-O(\eps)$ at binary rate
$\eps^2$ with high probability \cite[Theorem~3]{DoronMosheiffWootters2024}.
Their existence proof, however, leaves open how to construct such outer
codes explicitly. With this goal in mind, they asked what conditions on
a fixed outer code would suffice to ensure that, with high probability,
its concatenation with a single random binary linear inner code
approaches the GV bound at low rates
\cite[Question~1]{DoronMosheiffWootters2024}.

We address this question by extending their moment analysis to
random subcodes of a prescribed linear code after coordinate scaling.
The paper has two principal contributions.
\begin{enumerate}[leftmargin=2em]
\item We bound the probability of small minimum distance by two terms.
The first depends on the bias moments of the inner code. The second arises
from restricting the outer code to a scaled copy of the prescribed code $A$.
Averaging over the scalings expresses this term through the weight distribution
of $A^\perp$. A lower bound on the dual distance then suffices to control it.
\item For algebraic geometry (AG) codes on the Garcia--Stichtenoth
(GS) tower \cite{GarciaStichtenoth1996}, the Riemann--Roch bound on dual
distance makes the additional term exponentially small. We fix one inner
encoding map throughout the tower at each target rate. The resulting distance
lower bound holds with failure probability exponentially small in the block
length. Its coefficient of $\sqrt R$ agrees with that in the low-rate GV
expansion
\[
\delta_{\mathrm{GV}}(R)
=\frac12-\sqrt{\frac{\ln2}{2}}\sqrt R+O(R^{3/2})\qquad(R\to0).
\]
For each fixed target rate, the AG encoder and decoder also give polynomial-time
encoding and unique decoding up to the designed radius.
\end{enumerate}

\subsection{Main results}
\label{subsec:main-results}

\noindent\textbf{The codes considered in this paper.}
We study concatenated codes whose outer code is a random subcode of a scaled copy of a prescribed linear code and whose inner encoding map is the same at every coordinate.
Let $Q=2^{k_0}$, fix a binary $[n_0,k_0]$ linear code $\Cin$, and fix an $\Ftwo$-linear isomorphism $\phi:\FQ\to\Cin$.
Let $A\leq\FQ^n$ be a prescribed $K$-dimensional linear code, called the \emph{base code}, and fix an $\FQ$-linear encoding isomorphism $E_A:\FQ^K\to A$.
We select a $k$-dimensional subcode, where $1\leq k\leq K$.

Choose a random $k$-dimensional subspace $U\leq\FQ^K$ and, independently, choose $\alpha$ uniformly from $(\FQ^\times)^n$.
Let $\diag(\alpha)$ denote the diagonal matrix with entries $\alpha_1,\ldots,\alpha_n$. We call $A_\alpha:=\diag(\alpha)A$ the \emph{scaled base code} and define
\[
\Cout(\alpha,U):=\diag(\alpha)E_A(U)\subseteq A_\alpha,\qquad
C(\alpha,U):=\Cout(\alpha,U)\circ_\phi\Cin;
\]
see Section \ref{sec:prelim} for the precise definition of the concatenated codes.
Thus $C(\alpha,U)$ has block length $N=nn_0$ and binary dimension $kk_0$.
We choose the distribution of $U$ so that
\begin{equation}
\Prob_U[\ell|_U=0]\leq Q^{-k}
\quad\mathrm{for\ every\ nonzero\ }\FQ\text{-}\mathrm{linear\ functional\ }\ell:\FQ^K\to\FQ.
\label{H}
\end{equation}
The uniform distribution on $k$-dimensional subspaces satisfies \eqref{H}. \Cref{sec:subspace-distributions} gives further examples using graphs of linear maps.
In the moment expansion below, condition~\eqref{H} makes the average contribution from vectors outside $A_\alpha^\perp$ nonpositive.
Averaging over the coordinate scalings expresses the contribution from nonzero vectors in $A_\alpha^\perp$ in terms of the weight distribution of $A^\perp$ (\Cref{rem:scaling-role}).
These two steps are explained in \Cref{sec:transfer,sec:scaling}.
Moreover, the inclusion $\Cout(\alpha,U)\subseteq A_\alpha$ allows us to use the decoding algorithms available for $A$ (\Cref{subsec:inherited-decoding}).

\noindent\textbf{A general bound on the probability of small minimum distance.}
Write $\Tr:=\Tr_{\FQ/\Ftwo}$ and let $\omega_j\in\FQ$ be the unique element such that
\[
\phi(a)_j=\Tr(\omega_j a)
\qquad(a\in\FQ,\ 1\leq j\leq n_0).
\]
For an integer $r\geq1$, choose an ordered tuple
$V=((i_1,j_1),\ldots,(i_r,j_r))$ uniformly from
$\mathcal V_r:=(\{1,\ldots,n\}\times\{1,\ldots,n_0\})^r$, independently of $\alpha$.
Associate to $V$ the vector $g_V\in\FQ^n$ given by
\[
(g_V)_i:=\sum_{\ell:i_\ell=i}\omega_{j_\ell}
\qquad(1\leq i\leq n).
\]
For $0<\beta<1$ and an even integer $r\geq2$, define
\begin{equation}
\begin{aligned}
\mathsf T_0
&:=\frac{Q^k}{(\beta N)^r}
\#\{V\in\mathcal V_r:g_V=0\},\\
\mathsf T_A
&:=\frac{Q^k}{(\beta N)^r}
\E_\alpha\#\{V\in\mathcal V_r:0\ne g_V\in A_\alpha^\perp\}.
\end{aligned}
\label{eq:moment-contributions}
\end{equation}
We first prove the following bound.

\begin{theorem}[General probability bound]
\label{thm:intro-general}
For the construction above, suppose that the distribution of $U$ satisfies \eqref{H}.
For every $0<\beta<1$ and every even integer $r\geq2$,
\begin{equation}
\Prob_{\alpha,U}\!\left[\delta(C(\alpha,U))<\frac{1-\beta}{2}\right]
\leq\mathsf T_0+\mathsf T_A.
\label{eq:intro-general-bound}
\end{equation}
\end{theorem}

A binary linear code $C\leq\Ftwo^N$ is called \emph{$\beta$-balanced} if every nonzero $x\in C$ satisfies
$|\wt(x)/N-1/2|\leq\beta/2$, where $\wt(x)$ is its Hamming weight.
We will prove the following stronger estimate under the assumptions of
\Cref{thm:intro-general}:
\begin{equation}
\Prob_{\alpha,U}[C(\alpha,U)\ \mathrm{is\ not}\ \beta\text{-}\mathrm{balanced}]
\leq\mathsf T_0+\mathsf T_A.
\label{eq:intro-balanced-bound}
\end{equation}

When $A=\FQ^n$ and $U$ is a uniformly random $k$-dimensional subspace,
coordinate scaling preserves this distribution and $A^\perp=\{0\}$.
Thus $\mathsf T_A=0$ and only the tuples with $g_V=0$ enter the upper
bound, as in the random outer code argument of Doron, Mosheiff, and Wootters
\cite[Section~1.2]{DoronMosheiffWootters2024}.
With both constituent rates equal to $\eps$ and $\log_2 Q=\Omega(\eps^{-3})$,
this specialization recovers the $1/2-O(\eps)$ lower bound on relative
distance at binary rate $\eps^2$ in their Theorem~3
(see \Cref{subsec:random-outer-application}).
For a proper base code, \Cref{thm:intro-general} allows any subspace
distribution satisfying \eqref{H}, and the nonzero vectors in
$A_\alpha^\perp$ contribute the additional term $\mathsf T_A$.

The moment identity and the proof are given in \Cref{subsec:moment-identity,sec:transfer}.
In \Cref{sec:sufficient-conditions}, we construct distributions satisfying \eqref{H} and derive sufficient conditions for both terms to decay exponentially.

\noindent\textbf{Choice of  parameters.}
To obtain a useful distance guarantee, we choose the inner code and the base code so that both terms are small.
The first term, $\mathsf T_0$, is independent of $A$ and of the choice of the inner encoding map.
We choose an inner code satisfying the $\tau$-nice condition in \Cref{def:tau-nice}, and choose its rate and block length so that the bias of a uniformly random inner codeword has a sufficiently small moment generating function.
The estimates in \Cref{sec:zero-tuple-moments} then make $\mathsf T_0$ decay exponentially.

Our main technical contribution is an estimate for
$\mathsf T_A$, which arises from the restriction of the outer code
to the scaled base code $A_\alpha$.
In \Cref{sec:scaling}, we bound this additional term
using the weight distribution of $A^\perp$, and in
particular its minimum distance:
\[
\mathsf T_A
\leq\frac{Q^k}{\beta^r}(Q-1)^{1-d(A^\perp)}
\qquad(A^\perp\ne\{0\}).
\]
Thus, we choose $A$ with sufficiently large dual distance to make $\mathsf T_A$ decay exponentially as well.
The resulting conditions are stated in \Cref{cor:dual-distance-criterion}.

\noindent\textbf{A construction from AG codes.}
Let $A=C_L(D,G)$ be an AG code on a curve of genus $g$; see Section \ref{sec:garcia-stichtenoth} for the definition. The Riemann--Roch theorem gives $d(A^\perp)\geq\deg G-2g+2$.
On the GS tower, we can choose $\deg G$ so that this lower bound meets the requirement for $\mathsf T_A$, while $A$ has dimension large enough to contain the desired subcode.
Together with the choice of the inner code described above, this gives a concrete family to which \Cref{thm:intro-general} applies.
The precise construction is given in \Cref{subsec:gs-parameters,subsec:gs-construction} and yields the following result.

\begin{theorem}[GV asymptotics at low rates from AG codes (informal)]
\label{thm:intro-ag}
For every sufficiently small $\eps>0$, there is an explicit randomized
construction of binary concatenated codes $(C_m)_m$, with block lengths
$N_m$ and rates $R_m$, whose outer codes are subcodes of scaled AG codes.
For fixed $\eps$, the construction uses one inner encoding map,
independent of $m$, and the rates satisfy $\lim_{m\to\infty}R_m=\eps^2$.
Moreover, there is a constant $c>0$, independent of $\eps$ and $m$, such that,
for all sufficiently large $m$,
\[
\Prob_{\alpha_m,U_m}\!\left[\delta(C_m)\geq
h_2^{-1}(1-\eps^2-\eps^4)\right]
\geq1-2^{-c\eps^4N_m}.
\]
\end{theorem}
At the limiting rate $R=\lim_{m\to\infty}R_m=\eps^2$, the distance
target is $\delta_{\mathrm{GV}}(R+R^2)$, which differs from
$\delta_{\mathrm{GV}}(R)$ by $O(R^{3/2})$ as $R\to0$.
The full statement and parameter choices are given in
\Cref{thm:gs-gv,tab:gs-parameters}.
For each fixed $\eps$, we choose the inner code once by exhaustive
search and fix its encoding map. The probability in \Cref{thm:intro-ag}
is over the choice of the message subspace and the coordinate scalings.
With either subspace construction in \Cref{ex:graph-families}, the outer
seed has description length $O(\eps N_m)$ and the expected running time
of the construction is polynomial in $N_m$.

By \Cref{cor:fixed-outer-random-inner}, for each sufficiently small
$\eps>0$ and all sufficiently large $m$, there exists a subcode of a
scaled GS code whose concatenation with a uniformly random injective
$\Ftwo$-linear inner map meets the distance target of \Cref{thm:intro-ag}
with probability tending to one as $\eps\to0$.
This answers the fixed-outer-code question of Doron, Mosheiff, and Wootters \cite[Question~1]{DoronMosheiffWootters2024} for this family in the low-rate regime.

\subsection{Related work}

Guruswami and Indyk gave polynomial-time decoding algorithms for
low-rate concatenated codes that meet the GV bound using independently chosen
inner maps \cite{GuruswamiIndyk2004}.
Independent random inner encoding maps at all $n$ outer coordinates require separate random choices for each coordinate; specifying $n$ arbitrary binary linear $[n_0,k_0]$ inner maps directly requires $O(nn_0k_0)$ bits, compared with $O(n_0k_0)$ bits for a single reused map.

Barg, Justesen, and Thommesen studied concatenations of a fixed inner code with a random linear outer code using the weight distribution and cumulants of the inner code \cite{BargJustesenThommesen2001}.
In particular, their Proposition~3 relates the agreement of the resulting expansion at low rates with the GV expansion to the dual distance of the inner code.
Thus agreement with the leading GV coefficient in the setting of a random linear outer code is already known.
Their expansion keeps the inner code fixed as the total rate tends to zero.
In our GS family, the inner code is fixed along the tower at each target rate,
but its length grows as the target rate tends to zero.

Umeda, Sakakibara, and Kasahara used the average binary weight enumerator to
show that most binary images in their family of generalized AG codes meet the
GV bound \cite{UmedaSakakibaraKasahara1996}. In a binary expansion, each
symbol of $\FQ$ is represented by $k_0=\log_2 Q$ bits. In our construction,
each symbol is encoded by a word of length $n_0>k_0$ in a fixed $\tau$-nice
inner code. Ben-Aroya and Ta-Shma obtained explicit balanced binary codes
by concatenating Hermitian codes with the Hadamard code, whose nonzero
words all have the same weight \cite[Sections~3.2 and~3.3]{BenAroyaTaShma2013}.

Other recent approaches obtain capacity for list decoding by randomly deleting coordinates from AG codes over a fixed alphabet \cite{BrakensiekDharGopiZhang2025}, or take traces of AG codes to obtain binary codes \cite{CohenDoronGoldgraberManket2026}.
Here list decoding allows the decoder to return several candidate codewords; deleting coordinates is called puncturing.

\subsection*{Organization of the paper}
\Cref{sec:distance-estimates} introduces the construction and the moment identity, and proves \Cref{thm:intro-general}.
\Cref{sec:sufficient-conditions} gives distributions satisfying \eqref{H} and estimates the two terms in the bound.
\Cref{sec:ag-construction} derives the parameters, constructs AG base codes, and proves the distance theorem.
\Cref{sec:algorithms} treats randomness, encoding, and decoding.
\Cref{sec:limitations} compares the construction with random linear outer codes, explains the rate allocation, and discusses open problems.
\hyperref[app:poisson-bounds]{Appendix A} gives complementary bounds obtained by Poissonization.

\section{Construction framework and a general probability bound}
\label{sec:distance-estimates}
We recall the notation and the moment identity of Doron, Mosheiff, and Wootters, then use condition~\eqref{H} to prove \Cref{thm:intro-general}.

\subsection{Notation for codes and concatenation}
\label{sec:prelim}
For a positive integer $t$, let $[t]:=\{1,\ldots,t\}$.
We write $\ln$ for the natural logarithm and $\log_b$ for the logarithm to base $b$; an unsubscripted $\log$ also denotes the natural logarithm.
We use $O$ and $o$ with respect to the indicated limit.
In an inequality $f\leq g+O(h)$, with $h\geq0$, the notation means that
$f\leq g+Ch$ for some constant $C>0$ in the stated asymptotic regime.

A linear $[\ell,j]$ code over a field $F$ is a subspace $C\leq F^\ell$ of dimension $j$.
Its \emph{rate} is $j/\ell$.
The support of a word $c$ is $\supp(c):=\{i:c_i\ne0\}$, and its Hamming weight is $\wt(c):=|\supp(c)|$.
The dual code is $C^\perp:=\{y\in F^\ell:\sum_i c_i y_i=0\ \mathrm{for\ all}\ c\in C\}$.
The weight enumerator of $C$ is $W_C(x,y):=\sum_{c\in C}x^{\ell-\wt(c)}y^{\wt(c)}$.
The Hamming distance between two words of the same length is $d_H(x,y):=\wt(x-y)$.
The \emph{minimum Hamming distance} is $d(C):=\min\{\wt(c):0\neq c\in C\}$, with the convention $d(\{0\})=\infty$.
The \emph{relative minimum distance} is $\delta(C):=d(C)/\ell$.
We use the shorter term ``distance'' when the meaning is clear.

For a concatenated code, let $n$ be the outer block length, $n_0$ the binary inner block length, and $N:=nn_0$ the block length of the resulting binary code.
Let $\Cin\leq\Ftwo^{n_0}$ be the inner code of dimension $k_0$, and let $k$ be the dimension of the outer code.
Write $A\leq\FQ^n$ for the base code from which subcodes are selected, and let $K\geq k$ be its dimension.
Set $\rho_{\mathrm{in}}:=k_0/n_0$, $\rho_{\mathrm{out}}:=k/n$, $R:=kk_0/N=\rho_{\mathrm{in}}\rho_{\mathrm{out}}$, and $Q=2^{k_0}$.
Dimensions of codes over $\FQ$ and of binary codes are taken over $\FQ$ and $\Ftwo$, respectively.
For $S\subseteq[n]$, write $\FQ^S:=\{z\in\FQ^n:\supp(z)\subseteq S\}$.
The dual distance of $A$ is $d(A^\perp)$.

Write $\Tr:=\Tr_{\FQ/\Ftwo}$.
Fix a basis of $\FQ$ over $\Ftwo$.
For $C\leq\FQ^n$ and a fixed $\Ftwo$-linear isomorphism $\phi:\FQ\to\Cin$, denote the binary concatenation by
\[
C\circ_\phi\Cin
:=
\{
(\phi(c_1),\ldots,\phi(c_n))
:
c=(c_1,\ldots,c_n)\in C
\}
\leq\Ftwo^N.
\]
As the concatenated code depends on the choice of $\phi$, we indicate this dependence explicitly.

For a predicate $\mathsf P$, let $\1[\mathsf P]$ denote its indicator.
\begin{definition}
For any positive integer $\ell$ and $x\in\Ftwo^\ell$, define its unnormalized bias by $\operatorname{bias}(x):=\sum_{j=1}^\ell(-1)^{x_j}=\ell-2\wt(x)$.
For $x^{(1)},\ldots,x^{(n)}\in\Ftwo^{n_0}$, set
\[
\operatorname{bias}(x^{(1)},\ldots,x^{(n)})
:=\sum_{i=1}^n\operatorname{bias}(x^{(i)}).
\]
\end{definition}
This is the bias quantity used in \cite[Section~1.2]{DoronMosheiffWootters2024}, expressed in terms of a concatenated word.
For a concatenated word of length $N$ and $0<\beta<1$, the inequalities $\operatorname{bias}(x)>\beta N$ and $\wt(x)/N<(1-\beta)/2$ are equivalent.
This equivalence converts moment estimates into distance estimates.
A binary linear code $C\leq\Ftwo^N$ is called \emph{$\beta$-balanced} if, for every $0\ne x\in C$, $|\operatorname{bias}(x)|\leq\beta N$, or equivalently $(1-\beta)/2\leq\wt(x)/N\leq(1+\beta)/2$.

\noindent
\textbf{Our construction of concatenated codes.}
Now let $A\leq\FQ^n$ be the base code of dimension $K\geq k$, and fix an $\FQ$-linear encoding isomorphism $E_A:\FQ^K\to A$.
For $U$ sampled from a distribution $\mathcal U$ satisfying condition~\eqref{H} and $\alpha\in(\FQ^\times)^n$, define
\begin{equation}
  \Cout(\alpha,U):=\diag(\alpha)E_A(U)\subseteq A_\alpha:=\diag(\alpha)A.
  \label{eq:outer}
\end{equation}
See also Figure \ref{fig:construction} for the construction.
The dimension of $\Cout(\alpha,U)$ is exactly $k$, and $A_\alpha$ is its scaled base code.
When $Z$ denotes a random seed specifying the subspace, we write $U=U_Z$ and abbreviate $\Cout(\alpha,U_Z)$ as $\Cout(\alpha,Z)$.

\begin{figure}[tbp]
\centering
\[
\begin{array}{ccccc}
U & \xrightarrow{\ E_A\ } & E_A(U)
& \xrightarrow{\ \diag(\alpha)\ } & \Cout(\alpha,U)\\[-0.3ex]
\rotatebox{-90}{$\subseteq$} && \rotatebox{-90}{$\subseteq$} && \rotatebox{-90}{$\subseteq$}\\[-0.3ex]
\FQ^K & \xrightarrow{\ E_A\ } & A
& \xrightarrow{\ \diag(\alpha)\ } & A_\alpha:=\diag(\alpha)A
\end{array}
\]
\[
\Cout(\alpha,U)\xrightarrow{\ \phi^{\times n}\ }
C(\alpha,U)\subseteq A_\alpha\circ_\phi\Cin.
\]
\caption{A random subspace $U$ determines $E_A(U)\subseteq A$ in the fixed base code. Independent coordinate scaling gives the outer code inside the scaled base code $A_\alpha$. The map $\phi^{\times n}$ applies the same inner encoding map to every coordinate. }
\label{fig:construction}
\end{figure}

\subsection{A moment identity}
\label{subsec:moment-identity}
To analyze the distance of the concatenated code, we expand moments of the biases of its nonzero codewords.
The following identity holds for any fixed outer code, without assuming a probability distribution on that code.
Recall the trace representation $\phi(a)_j=\Tr(\omega_j a)$ used in \Cref{subsec:main-results}.
It exists because coordinate $j$ of $\phi$ is an $\Ftwo$-linear functional on $\FQ$; the nondegeneracy of the trace pairing gives the uniqueness of $\omega_j$.

For an integer $r\geq0$, let $\mathcal V_r:=([n]\times[n_0])^r$, with $\mathcal V_0$ consisting only of the empty tuple.
For $V=((i_1,j_1),\ldots,(i_r,j_r))\in\mathcal V_r$, define
\[
 (g_V)_i:=\sum_{\ell:i_\ell=i}\omega_{j_\ell}\in\FQ
 \qquad(i\in[n]).
\]
We have $|\mathcal V_r|=N^r$.
Permuting the outer coordinates induces a bijection on the tuples and the same permutation on the coordinates of $g_V$.
Define
\[
\mathcal Z_r:=\{V\in\mathcal V_r:g_V=0\},\qquad
p_r^{(0)}:=\frac{|\mathcal Z_r|}{N^r}
=\Prob_{V\in\mathcal V_r}[g_V=0],
\]
where $V$ is uniform on $\mathcal V_r$.
In particular, $p_0^{(0)}=1$.

Let $C\leq\FQ^n$ be a $k$-dimensional outer code, and fix an $\FQ$-linear encoding isomorphism $u\mapsto c(u)$ from $\FQ^k$ to $C$.
Set $X_u:=\operatorname{bias}(\phi(c(u)_1),\ldots,\phi(c(u)_n))$.
In what follows, $\E_{u\neq0}$ denotes the uniform average over $u\in\FQ^k\setminus\{0\}$.
\begin{lemma}[{Moment identity; cf.\ \cite[Section~1.2]{DoronMosheiffWootters2024}}]
\label{lem:concatenated-moment}
For every integer $r\geq1$, the following identity holds:
\begin{equation}
 \E_{u\neq0}X_u^r
 =\frac{1}{Q^k-1}\sum_{V\in\mathcal V_r}
   \left(Q^k\1[g_V\in C^\perp]-1\right).
 \label{eq:concatenated-moment}
\end{equation}
\end{lemma}
\begin{proof}
Expanding $X_u^r$ over $V\in\mathcal V_r$ and using the trace representation of $\phi$ gives $X_u^r=\sum_{V\in\mathcal V_r}(-1)^{\Tr(c(u)\cdot g_V)}$.
For each $V$, character orthogonality gives $Q^k\1[g_V\in C^\perp]$ for the sum over all messages. Removing the $u=0$ term, dividing by $Q^k-1$, and summing over $V$ gives \eqref{eq:concatenated-moment}.
\end{proof}

\subsection{Proof and consequences of the general bound}
\label{sec:transfer}
For uniformly random $\alpha\in(\FQ^\times)^n$, set $A_\alpha:=\diag(\alpha)A$.
To normalize the second count in \eqref{eq:moment-contributions}, define, for $r\geq1$,
\[
\mathcal L_r(A,\Cin):=\frac1{N^r}\E_\alpha\#\{V\in\mathcal V_r:g_V\neq0,\ g_V\in A_\alpha^\perp\}.
\]
Although $g_V$ is defined using $\phi$, this average is independent of $\phi$.
Set $\mathcal L_0(A,\Cin):=0$.
Both $p_r^{(0)}$ and $\mathcal L_r(A,\Cin)$ are normalized by the number $N^r$ of ordered tuples.
Since their defining events are disjoint, we have
\[
p_r^{(0)}+\mathcal L_r(A,\Cin)
=\Prob_{\alpha,V}[g_V\in A_\alpha^\perp],
\]
where $V$ is uniform on $\mathcal V_r$ and independent of $\alpha$.

Write $C:=\Cout(\alpha,U)\circ_\phi\Cin$.
The distance bound \eqref{eq:intro-general-bound} takes the form
\begin{equation}
\Prob_{\alpha,U}\!\left[\delta(C)<\frac{1-\beta}{2}\right]
\leq\frac{Q^k}{\beta^r}
\left(p_r^{(0)}+\mathcal L_r(A,\Cin)\right)
=\mathsf T_0+\mathsf T_A
\label{eq:transfer}
\end{equation}
for $0<\beta<1$ and even $r\geq2$.
Below, we prove the stronger estimate \eqref{eq:intro-balanced-bound}.
\begin{proof}[Proof of \Cref{thm:intro-general}]
For each nonzero message $u\in\FQ^k$ of the outer code, let $X_u$ denote the unnormalized bias of the corresponding binary codeword, and set $N_{\mathrm{bad}}:=\#\{u\in\FQ^k\setminus\{0\}:|X_u|>\beta N\}$.
Since $r$ is even, $\1[|X_u|>\beta N]\leq X_u^r/(\beta N)^r$, and hence
\begin{equation}
N_{\mathrm{bad}}\leq\frac{Q^k-1}{(\beta N)^r}\E_{u\neq0}X_u^r.
\label{eq:bad-message-moment-bound}
\end{equation}
Substituting \eqref{eq:concatenated-moment} in \eqref{eq:bad-message-moment-bound}
cancels the factor $Q^k-1$ and gives
\[
(\beta N)^r N_{\mathrm{bad}}
\leq \sum_{V\in\mathcal V_r}
\bigl(Q^k\1[g_V\in\Cout(\alpha,U)^\perp]-1\bigr).
\]
For a fixed tuple $V\in\mathcal V_r$ and scaling $\alpha$, define $\ell_{V,\alpha}(x):=\sum_{i=1}^n(g_V)_i\alpha_i(E_A(x))_i$.
By the definition of $\Cout(\alpha,U)$, the conditions $g_V\in\Cout(\alpha,U)^\perp$ and $\ell_{V,\alpha}|_U=0$ are equivalent.

We classify the tuples according to $g_V$.
First, if $g_V=0$, then $g_V\in\Cout(\alpha,U)^\perp$ for every $\alpha,U$,
since every linear code contains the zero vector.
These tuples therefore contribute $(Q^k-1)|\mathcal Z_r|$ to the sum above.

Next, suppose $g_V\neq0$.
If $g_V\notin A_\alpha^\perp$, then $\ell_{V,\alpha}$ is a nonzero linear functional.
Condition~\eqref{H} gives $\Prob_U[\ell_{V,\alpha}|_U=0]\leq Q^{-k}$, and thus $\E_U[Q^k\1[g_V\in\Cout(\alpha,U)^\perp]-1]\leq0$.
Consequently, these tuples make no positive contribution after averaging over $U$.

It remains to consider $g_V\neq0$ with $g_V\in A_\alpha^\perp$.
In this case, $\ell_{V,\alpha}=0$, so $g_V\in\Cout(\alpha,U)^\perp$ for every $U$.
Averaging first over $U$ and then over $\alpha$, the positive contribution from tuples with $g_V\ne0$ is at most $Q^k\sum_{V:g_V\neq0}\Prob_\alpha[g_V\in A_\alpha^\perp]$.

Combining these observations with \eqref{eq:bad-message-moment-bound} gives
\begin{equation}
\E_{\alpha,U}N_{\mathrm{bad}}
\leq\frac{Q^k}{\beta^r}\left(p_r^{(0)}+\mathcal L_r(A,\Cin)\right).
\label{eq:bad-count}
\end{equation}
The code $C$ fails to be $\beta$-balanced if and only if $N_{\mathrm{bad}}\geq1$.
Markov's inequality $\Prob[N_{\mathrm{bad}}\geq1]\leq\E N_{\mathrm{bad}}$ and \eqref{eq:bad-count} give the stronger estimate \eqref{eq:intro-balanced-bound}.
Since $\delta(C)<(1-\beta)/2$ implies that $C$ is not $\beta$-balanced, this also proves \eqref{eq:transfer} and \eqref{eq:intro-general-bound}.
\end{proof}

The moment identity follows the character orthogonality approach of Doron, Mosheiff, and Wootters \cite[Section~1.2]{DoronMosheiffWootters2024}.
The additional step is to separate the vectors in the dual of the scaled base code from those for which \eqref{H} gives a nonpositive average contribution.

\section{Sufficient conditions for exponential decay}
\label{sec:sufficient-conditions}
To apply \Cref{thm:intro-general}, the distribution of $U$ must satisfy
condition~\eqref{H}.
We construct such distributions in \Cref{sec:subspace-distributions}, then
estimate $\mathsf T_0$ and $\mathsf T_A$ in
\Cref{sec:zero-tuple-moments,sec:scaling}, respectively.
Exponential decay of both terms gives the desired bound on the failure probability.

\subsection{Distributions on subspaces}
\label{sec:subspace-distributions}
For $1\leq k\leq K$, write $\operatorname{Gr}_Q(k,K)$ for the set of all $k$-dimensional $\FQ$-subspaces of $\FQ^K$.
We first give a standard distribution satisfying \eqref{H}, then construct distributions that use fewer random elements.

The uniform distribution on $\operatorname{Gr}_Q(k,K)$ satisfies condition~\eqref{H}.
Indeed, counting the $k$-dimensional subspaces contained in the kernel of a nonzero linear functional $\ell$ gives
\[
\Prob_{U\sim\operatorname{Gr}_Q(k,K)}[\ell|_U=0]
 =\frac{Q^{K-k}-1}{Q^K-1}
 <Q^{-k}.
\]

Condition~\eqref{H} allows us to choose a restricted family of subspaces.
We use graphs of linear maps and give two constructions: one using Toeplitz matrices and one using multiplication in an extension field.
Set $s_0:=K-k$ and identify $\FQ^K$ with $\FQ^k\oplus\FQ^{s_0}$.
An arbitrary linear map from $\FQ^{s_0}$ to $\FQ^k$ requires $ks_0$ field elements to specify.
For $s_0>0$, a Toeplitz matrix uses only $k+s_0-1=K-1$ field elements.
The required property is that the image of each fixed nonzero input be uniform on $\FQ^k$.
As the next lemma shows, this property is sufficient for the distribution on the associated graphs to satisfy \eqref{H}.
Let $Z$ be a random seed and let $H_Z:\FQ^{s_0}\to\FQ^k$ be a random linear map such that
\begin{equation}
\Prob_Z[H_Z(b)=a]=Q^{-k}
\quad\mathrm{for\ every\ }0\ne b\in\FQ^{s_0}\ \mathrm{and\ every\ }a\in\FQ^k.
\label{eq:uniform-output}
\end{equation}
This is the uniformity property used in \cite[Definition~11]{BrehmIshaiResch2026}.
Define the graph of $H_Z^\top$ by
\[
U_Z:=\{(u,H_Z^\top u):u\in\FQ^k\}\leq\FQ^k\oplus\FQ^{s_0},
\]
where $u$ is a column vector.

\begin{lemma}
\label{lem:universal}
If $(H_Z)_Z$ satisfies \eqref{eq:uniform-output}, then the distribution of $U_Z$ satisfies condition~\eqref{H}.
More precisely, writing any nonzero linear map $\ell:\FQ^k\oplus\FQ^{s_0}\to\FQ$ as $\ell(u,v)=a\cdot u+b\cdot v$, we have
\[
 \Prob_Z[\ell|_{U_Z}=0]
 =\begin{cases}
    0,&b=0,\\
    Q^{-k},&b\neq0.
  \end{cases}
\]
\end{lemma}
\begin{proof}
On the graph, $\ell(u,H_Z^\top u)=(a+H_Z(b))\cdot u$.
The restriction vanishes identically if and only if $H_Z(b)=-a$.
If $b=0$, then $\ell\neq0$ implies $a\neq0$, so this event cannot occur.
If $b\neq0$, \eqref{eq:uniform-output} gives probability $Q^{-k}$.
\end{proof}

\begin{example}
\label{ex:graph-families}
When $s_0=0$, set $U_Z=\FQ^k$; no randomness is needed to specify the subspace $U_Z$.
In the following constructions, assume $s_0>0$.
\begin{enumerate}
\item \emph{Construction using Toeplitz matrices.}
Choose a uniformly random Toeplitz matrix $T\in\FQ^{s_0\times k}$, and set $H_Z(b)=T^\top b$.
Write $T_{i,j}=z_{j-i}$, and let $i_0$ be the largest index for which $b_{i_0}\ne0$.
Fix all diagonal variables except $z_{1-i_0},\ldots,z_{k-i_0}$.
The output $(T^\top b)_j$ contains $z_{j-i_0}$ with coefficient $b_{i_0}$, and every diagonal index appearing in the remaining terms is larger than $j-i_0$.
Thus, the linear part of the affine map from the remaining $k$ variables to $T^\top b$ is represented by a triangular matrix whose diagonal entries are all $b_{i_0}$, and is therefore invertible.
Consequently, $T^\top b$ is uniform on $\FQ^k$.
Toeplitz matrices give a standard example satisfying \eqref{eq:uniform-output}; see, for example, \cite{BrehmIshaiResch2026}.

\item \emph{Construction using multiplication in an extension field.}
Let $m_0=\max\{s_0,k\}$ and $\mathbb K=\mathbb F_{Q^{m_0}}$.
Fix an $\FQ$-linear injection $\iota:\FQ^{s_0}\hookrightarrow\mathbb K$ and an $\FQ$-linear surjection $P:\mathbb K\twoheadrightarrow\FQ^k$.
For uniformly random $z\in\mathbb K$, define $H_z(b)=P(z\,\iota(b))$.
If $b\neq0$, multiplication by $\iota(b)$ is a bijection of $\mathbb K$, so $H_z(b)$ is uniform.
\end{enumerate}
\end{example}

\subsection{\texorpdfstring{Controlling $\mathsf T_0$}{Controlling the first term}}
\label{sec:zero-tuple-moments}
To bound the first term $\mathsf T_0$ in \Cref{thm:intro-general}, we impose the following condition on the weight distribution of the dual of the inner code.
\begin{definition}[{\cite[Definition 2.2]{DoronMosheiffWootters2024}}]
\label{def:tau-nice}
Let $0<\tau<\rho_{\mathrm{in}}<1$.
A binary $[n_0,k_0]$ code $\Cin\leq\Ftwo^{n_0}$ of rate $\rho_{\mathrm{in}}=k_0/n_0$ is \emph{$\tau$-nice} if
\begin{equation}
 c_i:=\#\{x\in\Cin^\perp:\wt(x)=i\}
 \leq \binom{n_0}{i}2^{-n_0(\rho_{\mathrm{in}}-\tau)}
 \label{eq:nice}
\end{equation}
for every $1\leq i\leq n_0$.
\end{definition}
A uniformly random binary $[n_0,k_0]$ linear code is $\tau$-nice with
probability at least $1-n_0 2^{-\tau n_0}$.
Indeed, for every nonzero $x\in\Ftwo^{n_0}$,
\[
\Prob[x\in\Cin^\perp]
=\frac{2^{n_0-k_0}-1}{2^{n_0}-1}\leq2^{-k_0}.
\]
Hence $\E[c_i]\leq\binom{n_0}{i}2^{-k_0}$ for $1\leq i\leq n_0$.
Markov's inequality bounds the probability of violating \eqref{eq:nice}
at any fixed weight by $2^{-\tau n_0}$; a union bound gives the assertion.

Neither the moment identity nor the general probability bound requires \eqref{eq:nice}.
Condition \eqref{eq:nice} is used only when bounding the term arising from the inner code, starting with \Cref{prop:tau-mgf}.

Bounding $\mathsf T_0$ reduces to estimating the number $|\mathcal Z_r|$ of tuples satisfying $g_V=0$.
We estimate this number using the moment generating function (MGF) of the bias of a uniformly random word of $\Cin$.

Let $\mathsf U$ be uniform on $\FQ$, and set $Y:=\operatorname{bias}(\phi(\mathsf U))$.
Its moment generating function is
\begin{equation}
\label{eq:inner-bias-mgf}
\mathsf M_{\Cin}(s)
:=\E[e^{sY}]
=\frac1Q\sum_{a\in\FQ}\exp(s\operatorname{bias}(\phi(a))).
\end{equation}
The function $\ln\mathsf M_{\Cin}$ is the cumulant generating function used in the analysis of a fixed inner code and a random outer code by Barg, Justesen, and Thommesen \cite{BargJustesenThommesen2001}, expressed in terms of the bias variable.

Let $\mathsf U_1,\ldots,\mathsf U_n$ be independent uniform $\FQ$-valued random variables, and set $Y_i:=\operatorname{bias}(\phi(\mathsf U_i))$ and $S_n:=\sum_{i=1}^nY_i$.

\begin{lemma}
\label{lem:zero-tuple-mgf}
For every integer $r\geq1$, we have $|\mathcal Z_r|=\E[S_n^r]$.
Consequently, for every even $r\geq2$ and every $s>0$,
\begin{equation}
|\mathcal Z_r|
\leq\frac{r!}{2s^r}
\left(\mathsf M_{\Cin}(s)^n+\mathsf M_{\Cin}(-s)^n\right).
\label{eq:Zr-mgf-general}
\end{equation}
\end{lemma}
\begin{proof}
Using
\[
\operatorname{bias}(\phi(\mathsf U_i))
=\sum_{j=1}^{n_0}(-1)^{\Tr(\omega_j\mathsf U_i)},
\]
we expand $S_n^r$ over the ordered tuples $V=((i_1,j_1),\ldots,(i_r,j_r))\in\mathcal V_r$.
Independence gives
\[
\E[S_n^r]
=\sum_{V\in\mathcal V_r}\prod_{i=1}^n
\E_{\mathsf U_i}\left[(-1)^{\Tr(\mathsf U_i(g_V)_i)}\right].
\]
By character orthogonality, the product equals one when $g_V=0$ and zero otherwise.
Hence $|\mathcal Z_r|=\E[S_n^r]$.
If $r$ is even, then $\cosh(sx)\geq(sx)^r/r!$ for every real $x$, so
\[
\E[S_n^r]
\leq\frac{r!}{s^r}\E[\cosh(sS_n)]
=\frac{r!}{2s^r}
\left(\mathsf M_{\Cin}(s)^n+\mathsf M_{\Cin}(-s)^n\right).
\]
\end{proof}

Set $\Delta:=\rho_{\mathrm{in}}-\tau>0$, and for $s\geq0$ define
\[
\Gamma_{\Delta,n_0}(s)
:=1+2^{-n_0\Delta}\left((1+\tanh s)^{n_0}-1\right).
\]

\begin{proposition}
\label{prop:tau-mgf}
Suppose that $\Cin$ has rate $\rho_{\mathrm{in}}$ and is $\tau$-nice.
For every even $r\geq2$ and every $s>0$,
\begin{equation}
|\mathcal Z_r|
\leq\frac{r!}{s^r}(\cosh s)^N
\Gamma_{\Delta,n_0}(s)^{N/n_0}.
\label{eq:Zr-mgf-tau-nice}
\end{equation}
\end{proposition}
\begin{proof}
We first bound the moment generating function $\mathsf M_{\Cin}$ in \eqref{eq:inner-bias-mgf}.
For $s\geq0$, the binary MacWilliams identity \cite[Chapter~5]{MacWilliamsSloane1977} gives
\begin{equation}
\mathsf M_{\Cin}(s)
=(\cosh s)^{n_0}\sum_{z\in\Cin^\perp}(\tanh s)^{\wt(z)}.
\label{eq:inner-mgf-macwilliams}
\end{equation}
Since $\Cin$ is $\tau$-nice, for every $j\geq1$,
\[
c_j=\#\{z\in\Cin^\perp:\wt(z)=j\}
\leq2^{-n_0\Delta}\binom{n_0}{j}.
\]
The binomial theorem therefore yields
\[
\sum_{z\in\Cin^\perp}(\tanh s)^{\wt(z)}
\leq1+2^{-n_0\Delta}\bigl((1+\tanh s)^{n_0}-1\bigr)
=\Gamma_{\Delta,n_0}(s).
\]
For $-s$, the expansion contains terms of alternating signs; taking their absolute values gives the same upper bound.
Thus $\mathsf M_{\Cin}(\pm s)\leq(\cosh s)^{n_0}\Gamma_{\Delta,n_0}(s)$.
Substitution into \Cref{lem:zero-tuple-mgf} gives
\[
|\mathcal Z_r|
\leq\frac{r!}{s^r}(\cosh s)^N
\Gamma_{\Delta,n_0}(s)^{N/n_0}.
\]
\end{proof}

We will use the following elementary estimate:
\begin{equation}
\Gamma_{\Delta,n_0}(s)
\leq1+2^{-n_0\Delta}(1+\tanh s)^{n_0}
=1+\exp\!\left[-n_0\bigl(\Delta\ln2-\ln(1+\tanh s)\bigr)\right].
\label{eq:Gamma-general}
\end{equation}
Writing $\kappa(s):=\Delta\ln2-\ln(1+\tanh s)$, we have $0\leq\Gamma_{\Delta,n_0}(s)-1\leq e^{-n_0\kappa(s)}$.
Thus, even when the parameters vary, $n_0\kappa(s)\to\infty$ implies $\Gamma_{\Delta,n_0}(s)\to1$.
If, in addition, $\kappa(s)\geq\kappa_0>0$ holds uniformly, then $\Gamma_{\Delta,n_0}(s)-1$ is exponentially small in $n_0$.
In the applications at low rates below, we estimate $n_0^{-1}\ln\Gamma_{\Delta,n_0}(s)$ directly to the required precision.

Since $p_r^{(0)}=|\mathcal Z_r|/N^r$, \Cref{prop:tau-mgf} gives the following bound whenever $\Cin$ is $\tau$-nice, $s>0$, and $r\geq2$ is even:
\[
\mathsf T_0
=\frac{Q^k}{\beta^r}p_r^{(0)}
\leq
\overline{\mathsf T}_0
:=
\frac{Q^k r!}{(\beta sN)^r}
(\cosh s)^N
\Gamma_{\Delta,n_0}(s)^{N/n_0}.
\]
Fix $s>0$ and $0<\beta<1$. For even $r$ with $r/N\to\lambda>0$,
Stirling's formula gives
\[
\frac1N\ln\frac{r!}{(\beta sN)^r}
=\lambda\ln\frac{\lambda}{\beta s}-\lambda+o_N(1).
\]
The limiting expression is minimized at $\lambda=\beta s$.
The factorial ratio and $(\cosh s)^N$ then contribute $\ln\cosh s-\beta s$
to the limiting exponent. The derivative of this expression is $\tanh s-\beta$,
so we choose
\begin{equation}
s_\beta:=\operatorname{arctanh}(\beta),\qquad \lambda_\beta:=\beta s_\beta.
\label{eq:gv-saddlepoint}
\end{equation}
Define
\begin{equation}
\mathsf I(\beta):=\beta s_\beta-\ln\cosh(s_\beta)=\ln2\left(1-h_2\!\left(\frac{1-\beta}{2}\right)\right).
\label{eq:binary-large-deviation-rate}
\end{equation}
This is the rate function for large deviations of a uniform $\{-1,1\}$-valued random variable, and it yields the exponent associated with the GV bound in the analysis at low rates below.

Since $\tanh(s_\beta)=\beta$, we also have
\begin{equation}
\Gamma_{\Delta,n_0}(s_\beta)\leq1+\exp\!\left[-n_0\bigl(\Delta\ln2-\ln(1+\beta)\bigr)\right].
\label{eq:Gamma-at-saddle}
\end{equation}
Thus, at $s=s_\beta$, the factor $\Gamma_{\Delta,n_0}(s_\beta)$ is determined by the size of $n_0\bigl(\Delta\ln2-\ln(1+\beta)\bigr)$.
Taking even $r=r_N$ with $r/N\to\lambda_\beta$ gives
\begin{equation}
\frac1N\ln\mathsf T_0
\leq\frac1N\ln\overline{\mathsf T}_0
=R\ln2-\mathsf I(\beta)
+\frac1{n_0}\ln\Gamma_{\Delta,n_0}(s_\beta)+o_N(1).
\label{eq:T0-exponent-general}
\end{equation}
In particular, $\mathsf T_0$ decays exponentially in $N$ if there is a constant $a_0>0$, independent of $N$, such that
\begin{equation}
\mathsf I(\beta)-R\ln2-\frac1{n_0}\ln\Gamma_{\Delta,n_0}(s_\beta)\geq a_0
\label{eq:zero-exponent-condition}
\end{equation}
for all sufficiently large $N$.

\subsection{\texorpdfstring{Controlling $\mathsf T_A$}{Controlling the second term}}
\label{sec:scaling}
To bound $\mathsf T_A$, fix a tuple $V$ with $g_V\ne0$.
The event $g_V\in A_\alpha^\perp$ is equivalent to $\diag(\alpha)g_V\in A^\perp$.
Independent scaling makes $\diag(\alpha)g_V$ uniform among vectors with the
same support as $g_V$. Its probability of belonging to $A^\perp$ is therefore
determined by the number of dual codewords with that support.

For a fixed moment order $r\geq1$ and a nonempty set $S\subseteq[n]$, let
\[
 N_A^\perp(S):=\#\{z\in A^\perp:\supp(z)=S\},
 \qquad
 M_S:=\#\{V\in\mathcal V_r:\supp(g_V)=S\}.
\]

For $1\leq w\leq n$, define $\pi_r(w):=N^{-r}\#\{V\in\mathcal V_r:\wt(g_V)=w\}$ and
\[
\theta_w(A):=\frac{1}{\binom nw}
\sum_{\substack{S\subseteq[n]\\|S|=w}}\frac{N_A^\perp(S)}{(Q-1)^w}.
\]
The quantity $\pi_r(w)$ is the proportion of tuples for which $\wt(g_V)=w$.
The quantity $\theta_w(A)$ is the probability that a uniformly random vector of weight $w$, obtained by first choosing its support uniformly and then choosing each nonzero entry uniformly, belongs to $A^\perp$.

Set $\Theta_A(Q):=\max_{1\leq w\leq n}\theta_w(A)$.
Writing $B_w^\perp:=\#\{z\in A^\perp:\wt(z)=w\}$ and partitioning by support gives
\[
\theta_w(A)=\frac{B_w^\perp}{\binom nw(Q-1)^w}.
\]
Thus, $\Theta_A(Q)$ is the maximum over $w$ of the proportion $\theta_w(A)$ of vectors of weight $w$ that belong to $A^\perp$, and is determined by the ordinary weight enumerator.
We use the numbers of codewords with prescribed supports to estimate this proportion.

\begin{theorem}
\label{thm:moment-counts}
For every integer $r\geq1$,
\[
\mathcal L_r(A,\Cin)=\sum_{w=1}^n\pi_r(w)\theta_w(A)
\leq\Theta_A(Q).
\]
\end{theorem}
\begin{proof}
Fix $0\neq g\in\FQ^n$, and let $S=\supp(g)$ be its support and $w=|S|$ its weight.
For uniformly random $\alpha\in(\FQ^\times)^n$, the vector $\diag(\alpha)g$ is uniform over the $(Q-1)^w$ vectors with support exactly $S$.
Since $\diag(\alpha)g\in A^\perp$ is equivalent to $g\in A_\alpha^\perp$, we have
\[
 \Prob_\alpha[g\in A_\alpha^\perp]
 =\frac{N_A^\perp(S)}{(Q-1)^w}.
\]

Summing this identity over $V$ gives
\[
 \E_\alpha\#\{V\in\mathcal V_r:g_V\neq0,\ g_V\in A_\alpha^\perp\}
 =\sum_{\varnothing\neq S\subseteq[n]}
 M_S\frac{N_A^\perp(S)}{(Q-1)^{|S|}}.
\]
For a permutation $\sigma$ of $[n]$ and $V=((i_1,j_1),\ldots,(i_r,j_r))\in\mathcal V_r$, define
\[
\sigma V:=((\sigma(i_1),j_1),\ldots,(\sigma(i_r),j_r)).
\]
The map $V\mapsto\sigma V$ is a bijection of $\mathcal V_r$, and the definition of $g_V$ gives
\[
(g_{\sigma V})_{\sigma(i)}=(g_V)_i\quad(i\in[n]),
\qquad
\supp(g_{\sigma V})=\sigma(\supp(g_V)).
\]
Thus, the uniform distribution on $\mathcal V_r$ is invariant under permutations of the outer coordinates, and $M_S$ depends only on $|S|$.
Denote the corresponding value by $M_w$.
Grouping by $w=|S|$ yields
\begin{equation}
\label{eq:Lr_develop}
\mathcal L_r(A,\Cin)
=\frac1{N^r}\sum_{w=1}^nM_w
\sum_{\substack{S\subseteq[n]\\|S|=w}}
\frac{N_A^\perp(S)}{(Q-1)^w}.
\end{equation}
On the other hand, the definition of $\pi_r(w)$ gives
\[
\pi_r(w)=\frac1{N^r}\sum_{\substack{S\subseteq[n]\\|S|=w}}M_S
=\frac{\binom nw M_w}{N^r}.
\]
Substituting this into \eqref{eq:Lr_develop} yields $\mathcal L_r(A,\Cin)=\sum_{w=1}^n\pi_r(w)\theta_w(A)$.
Since $\sum_{w=1}^n\pi_r(w)$ is the proportion of tuples for which $g_V\ne0$, we have $\sum_{w=1}^n\pi_r(w)=1-|\mathcal Z_r|/N^r\leq1$.
The final inequality follows from $\theta_w(A)\leq\Theta_A(Q)$ for every $w$.
\end{proof}
By nondegeneracy of the trace pairing, $(g_V)_i=0$ if and only if
$\sum_{\ell:i_\ell=i}x_{j_\ell}=0$ for every $x\in\Cin$.
Thus $\supp(g_V)$, and hence both $|\mathcal Z_r|$ and
$\mathcal L_r(A,\Cin)$ by the formula above, are independent of $\phi$.

\begin{remark}[The role of coordinate scaling]
\label{rem:scaling-role}
The decomposition in \Cref{thm:intro-general} also holds for a fixed $\alpha$, with the corresponding probability of $0\ne g_V\in A_\alpha^\perp$ in place of its average over $\alpha$.
The reduction to the dual weight distribution in \Cref{thm:moment-counts}, however, uses the independent uniform scalings.
For fixed $0\ne g$ with support $S$, they make $\diag(\alpha)g$ uniform among the vectors with support exactly $S$.
Without this averaging, the probability of dual membership can depend on the actual nonzero entries of $g_V$ and on the chosen inner encoding map.

For example, take $Q=8$, $n=2$, $\Cin=\Ftwo^3$, and
$A=\{(a,a):a\in\mathbb F_8\}$.
Then $A^\perp=A$, $d(A^\perp)=2$, and $\Theta_A(8)=1/7$.
For any isomorphism $\phi:\mathbb F_8\to\Ftwo^3$, the three trace coefficients $\omega_1,\omega_2,\omega_3$ form a basis over $\Ftwo$.
When $r=2$, exactly six of the $36$ ordered tuples give a nonzero vector in $A^\perp$: the two outer coordinates must be distinct and the inner indices must agree.
Thus
\[
\Prob_V[0\ne g_V\in A^\perp]=\frac16>\Theta_A(8)=\frac17.
\]
After independent coordinate scaling, the probability is
$\mathcal L_2(A,\Cin)=\pi_2(2)\theta_2(A)=1/14$.
This example shows that the upper bound by $\Theta_A(Q)$ need not hold when the scalings are omitted.
\end{remark}

\begin{lemma}[A uniform bound in terms of the dual distance]
\label{lem:dual-distance-density}
Suppose $A^\perp\neq\{0\}$, and set $d^\perp:=d(A^\perp)$.
For every nonempty set $S\subseteq[n]$, with $w:=|S|$,
\[
N_A^\perp(S)\leq
\begin{cases}
0,&w<d^\perp,\\
(Q-1)^{w-d^\perp+1},&w\geq d^\perp.
\end{cases}
\]
Consequently, for every even $r\geq2$,
\[
\mathcal L_r(A,\Cin)
\leq\Theta_A(Q)
\leq(Q-1)^{1-d^\perp}.
\]
If $A^\perp=\{0\}$, then $\mathcal L_r(A,\Cin)=\Theta_A(Q)=0$.
\end{lemma}

\begin{proof}
The case $w<d^\perp$ follows from the definition of minimum distance.
For $w\geq d^\perp$, remove any $d^\perp-1$ coordinates from $S$ and let $T$ be the remaining set of coordinates.
If two dual codewords with support exactly $S$ agree on $T$, their difference is supported on at most $d^\perp-1$ coordinates.
This difference also belongs to $A^\perp$, so it is zero by the definition of minimum distance.
Projection onto $T$ is therefore injective.
Each codeword is nonzero in every coordinate of $T$, so there are at most $(Q-1)^{|T|}=(Q-1)^{w-d^\perp+1}$ dual codewords with support $S$.
Dividing by $(Q-1)^w$ and averaging over supports gives $\theta_w(A)\leq(Q-1)^{1-d^\perp}$.
The conclusion follows from the definition of $\Theta_A(Q)$ and \Cref{thm:moment-counts}.
\end{proof}

The bound in \Cref{lem:dual-distance-density} is attained by any linear $[n,K]$ \emph{maximum distance separable (MDS)} code $A$ with $1\leq K<n$.
Such a code satisfies $d(A)=n-K+1$, and $A^\perp$ is MDS with $d(A^\perp)=K+1$.
For every $S$ of size $K+1$, the shortened dual code $A^\perp\cap\FQ^S$ has dimension one, and every nonzero word in it has support $S$.
Thus $N_A^\perp(S)=Q-1$ and $\theta_{K+1}(A)=(Q-1)^{-K}$, which gives $\Theta_A(Q)=(Q-1)^{-K}$.
If $K=n$, then $A^\perp=\{0\}$ and $\Theta_A(Q)=0$.

We now combine the estimates for the two terms.
If $A^\perp\ne\{0\}$, \Cref{lem:dual-distance-density} gives
\begin{equation}
\frac1N\log_2\mathsf T_A
\leq R+\frac rN\log_2\frac1\beta
-\rho_{\mathrm{in}}\frac{d(A^\perp)-1}{n}\log_Q(Q-1),
\label{eq:TA-exponent-general}
\end{equation}
with the convention $\log_2 0=-\infty$.

Under the assumptions of \Cref{thm:intro-general,prop:tau-mgf}, fix $0<\beta<1$ and choose even integers $r=r_N$ with $r/N=\lambda_\beta+o_N(1)$.

\begin{corollary}[A sufficient condition using the dual distance]
\label{cor:dual-distance-criterion}
For these choices, suppose that, for constants $a_0,a_A>0$ independent of $N$,
\begin{align*}
\mathsf I(\beta)-R\ln2-\frac1{n_0}\ln\Gamma_{\Delta,n_0}(s_\beta)&\geq a_0,\\
\rho_{\mathrm{in}}\frac{d(A^\perp)-1}{n}\log_Q(Q-1)
-R-\frac rN\log_2\frac1\beta&\geq a_A
\end{align*}
for all sufficiently large $N$.
The second condition is unnecessary when $A^\perp=\{0\}$.
Then there is $c>0$, independent of $N$, such that
\[
\Prob_{\alpha,U}\!\left[C(\alpha,U)\ \mathrm{is\ not}\ \beta\text{-}\mathrm{balanced}\right]
\leq e^{-cN}
\]
for all sufficiently large $N$.
\end{corollary}
\begin{proof}
Equations \eqref{eq:T0-exponent-general} and \eqref{eq:TA-exponent-general} make both exponents negative by constants independent of $N$.
Apply \eqref{eq:intro-balanced-bound}, absorbing the sum of the two exponential bounds into a smaller exponent constant.
\end{proof}

These conditions guide the parameter choices in the next section.
The exact weighted sum in \Cref{thm:moment-counts} may replace the estimate using the dual distance when more information about $A^\perp$ is available.
\hyperref[app:poisson-bounds]{Appendix A} gives further bounds obtained by Poissonization.
The effect of the allocation of the inner and outer rates is discussed in \Cref{subsec:rate-allocation}.

\section{Construction using AG codes}
\label{sec:ag-construction}
We apply the general bound to AG codes on the GS tower.
We first derive the parameter choices from the two exponent conditions.
We then construct the AG base codes $A$, scale their coordinates to obtain
$A_\alpha$, and choose random subcodes of $A_\alpha$ as the outer codes.

\subsection{Choice of the inner code and the required dual distance}
\label{subsec:gs-parameters}
Fix constants
\begin{equation}
c_0>2,\qquad c_A>\sqrt2(\ln2)^{3/2}.
\label{eq:gs-ca-threshold}
\end{equation}
For sufficiently small $\eps>0$, our target rate is $\eps^2$ and our distance target is
\begin{equation}
\delta_\eps:=h_2^{-1}(1-\eps^2-\eps^4),
\qquad \beta_\eps:=1-2\delta_\eps.
\label{eq:gs-target-distance}
\end{equation}
Thus $\mathsf I(\beta_\eps)=(\eps^2+\eps^4)\ln2$.
The margin above the target rate in the first exponent is $\eps^4\ln2$.
We first let $m\to\infty$ with $\eps$ fixed, and then let $\eps\to0$.
The notation $o_m(1)$ and $o_\eps(1)$ refers to these respective limits.

To control $\mathsf T_0$, choose the target inner rate slightly above
$\log_2(1+\beta_\eps)+\tau$, as suggested by \eqref{eq:Gamma-at-saddle}.
Here $\tau$ is the tolerance in the condition on the dual weight distribution
in \eqref{eq:nice}.
Taking $\tau=o(\eps)$ preserves the leading coefficient of the inner rate,
since $\log_2(1+\beta_\eps)=\Theta(\eps)$.
For a fixed inner rate, a smaller $\tau$ strengthens the moment bound, but
requires a larger inner block length to make $n_0 2^{-\tau n_0}$ small.
To see how these requirements fit together, temporarily leave $\tau$ unspecified
and set $\bar\rho_{\mathrm{in}}=\log_2(1+\beta_\eps)+2\tau$ and
$n_0=\lceil c_0\tau^{-1}\log_2(1/\eps)\rceil$.
With the rounding of $k_0$ described below, the sampling estimate and
\eqref{eq:Gamma-at-saddle} give
\[
n_0 2^{-\tau n_0}
=O\!\left(\frac{\eps^{c_0}\log(1/\eps)}{\tau}\right),
\qquad
\frac1{n_0}\ln\Gamma_{\rho_{\mathrm{in}}-\tau,n_0}(s_{\beta_\eps})
=O\!\left(\frac{\tau\eps^{c_0}}{\log(1/\eps)}\right).
\]
Making the first quantity tend to zero ensures that a randomly chosen inner
code is $\tau$-nice with probability tending to one.
We require the second to be $o(\eps^4)$ to preserve the margin in the exponent
of $\mathsf T_0$.
The choice $\tau=\eps^2$ satisfies both requirements for every $c_0>2$ and
also gives $\tau=o(\eps)$.
We therefore take
\begin{equation}
\tau:=\eps^2,\qquad
n_0:=\left\lceil c_0\eps^{-2}\log_2(1/\eps)\right\rceil,
\qquad
\bar\rho_{\mathrm{in}}:=\log_2(1+\beta_\eps)+2\eps^2.
\label{eq:inner-design}
\end{equation}
Choose an even integer $k_0$ with $|k_0-\bar\rho_{\mathrm{in}}n_0|\leq2$,
and set $\rho_{\mathrm{in}}:=k_0/n_0$ and $Q:=2^{k_0}$.
For sufficiently small $\eps$, we have $1\leq k_0<n_0$.
The parity condition makes $Q$ a square, as required for the GS tower.
The factor $\log_2(1/\eps)$ in $n_0$ makes the probability of failing the
$\tau$-nice condition tend to zero:
\begin{equation}
n_0 2^{-\tau n_0}
=O\!\left(\eps^{c_0-2}\log(1/\eps)\right)=o_\eps(1).
\label{eq:tau-sampling-failure}
\end{equation}
Hence a $\tau$-nice inner code exists for every sufficiently small $\eps$.
Fix any such code $\Cin$ and any $\Ftwo$-linear isomorphism $\phi:\FQ\to\Cin$.

We will choose the outer dimension using the actual rate $\rho_{\mathrm{in}}$
so that the binary rate never exceeds $\eps^2$.
This avoids a positive error from rounding $k_0$ in the rate term of
\eqref{eq:T0-exponent-general}.
The remaining correction $n_0^{-1}\ln\Gamma$ is
$O(\eps^{c_0+2}/\log(1/\eps))=o_\eps(\eps^4)$, as proved in
\Cref{lem:gs-exponents}.

For the second term, take $r/N=\lambda_{\beta_\eps}+o_N(1)$.
The entropy expansion and \eqref{eq:inner-design} give
\[
\beta_\eps=\sqrt{2\ln2}\,\eps+O(\eps^3),\qquad
\rho_{\mathrm{in}}=\sqrt{\frac2{\ln2}}\,\eps+O(\eps^2),\qquad
\lambda_{\beta_\eps}=2\ln2\,\eps^2+O(\eps^4).
\]
The positive contribution $(r/N)\log_2(1/\beta_\eps)$ in
\eqref{eq:TA-exponent-general} therefore has order
$\eps^2\log_2(1/\eps)$.
Since $\rho_{\mathrm{in}}$ has order $\eps$, a relative dual distance
of order $\eps\log_2(1/\eps)$ can offset this contribution.
We will realize it with AG base codes whose limiting rate is
\begin{equation}
R_A:=c_A\eps\log_2(1/\eps).
\label{eq:gs-base-rate}
\end{equation}
The condition on $c_A$ in \eqref{eq:gs-ca-threshold} makes
$2\ln2-c_A\sqrt{2/\ln2}$ negative.
The genus term in the dual distance will be negligible for our choice of $Q$.

\subsection{AG evaluation codes and the GS tower}
\label{sec:garcia-stichtenoth}

\begin{definition}[{\cite[Chapter~2, Sections~2 and~4]{Stichtenoth2009}}]
Let $X/\FQ$ be a smooth projective geometrically irreducible curve of genus $g$.
Let $D=P_1+\cdots+P_n$ be a sum of distinct $\FQ$-rational points, and let $G$ be an $\FQ$-rational divisor whose support is disjoint from that of $D$, with $\deg G=\mu<n$.
Set
\[
L(G):=\{f\in\FQ(X):\operatorname{div}(f)+G\geq0\}\cup\{0\}.
\]
The associated AG evaluation code is defined by
\[
C_L(D,G):=\{(f(P_1),\ldots,f(P_n)):f\in L(G)\}.
\]
\label{def:evaluation-code}
\end{definition}
Write $A:=C_L(D,G)$ for the base code/
The resulting code satisfies $d(A)\geq n-\mu$.
If $\mu\geq2g-1$, the Riemann--Roch theorem gives $K=\dim A=\mu+1-g$ \cite[Section~1.5]{Stichtenoth2009}.

For $2g-1\leq\mu<n$, the standard dual distance bound for AG codes gives
$d(A^\perp)\geq d_0:=\mu-2g+2$ \cite[Chapter~2]{Stichtenoth2009}.
By \Cref{lem:dual-distance-density,thm:moment-counts}, it follows that
\[
\mathcal L_r(A,\Cin)\leq\Theta_A(Q)\leq(Q-1)^{1-d_0},\qquad
\mathsf T_A\leq\frac{Q^k}{\beta^r}(Q-1)^{1-d_0}.
\]
Thus the degree of $G$ and the genus of the curve directly control $\mathsf T_A$.

\medskip\noindent\textbf{The GS tower.}
Let $Q=q^2$ be a square prime power.
We use the second GS tower \cite{GarciaStichtenoth1996}, whose function fields are defined by
\[
F_1=\mathbb F_Q(x_1),\qquad
F_{m+1}=F_m(x_{m+1}),\qquad
x_{m+1}^{q}+x_{m+1}=\frac{x_m^{q}}{x_m^{q-1}+1}.
\]
Let $\mathcal X_m$ be the smooth projective curve with function field $F_m$
and genus $g_m$. We identify its closed points with the places of $F_m$.
Write $P_{\infty,m}$ for the unique point above the pole of $x_1$.
Shum et al.\ used this tower to compute bases of Riemann--Roch spaces \cite{ShumEtAl2001}.
For every $a\in\mathbb F_Q$ with $a^q+a\ne0$, the rational place $x_1=a$
of $F_1$ splits completely in $F_m$.
We take all the points above these $q^2-q$ places as evaluation points
$P_{1,m},\ldots,P_{n_m,m}$, so $n_m=(q^2-q)q^{m-1}$.
These evaluation points yield the optimal asymptotic ratio \cite{GarciaStichtenoth1996,ShumEtAl2001}:
\[
\frac{g_m}{n_m}=\gamma_Q+o_m(1),
\qquad
\gamma_Q:=\frac1{\sqrt Q-1}.
\]

For a target rate $R_A$ of the AG base code satisfying $\gamma_Q<R_A<1-\gamma_Q$, set $\mu_m:=\lfloor R_A n_m+g_m-1\rfloor$, $G_m:=\mu_mP_{\infty,m}$, $D_m:=P_{1,m}+\cdots+P_{n_m,m}$, and $A_m:=C_L(D_m,G_m)$.
Also let $K_m:=\dim_{\FQ}A_m$ and $d_{0,m}:=\mu_m-2g_m+2$.
In the notation of \Cref{thm:intro-general}, we take $A=A_m$.
For all sufficiently large $m$, we have $2g_m-1\leq\mu_m<n_m$, and the Riemann--Roch theorem together with $g_m/n_m=\gamma_Q+o_m(1)$ gives
\begin{equation}
\frac{K_m}{n_m}=R_A+o_m(1),
\qquad
\frac{d_{0,m}}{n_m}=R_A-\gamma_Q+o_m(1).
\label{eq:gs-dim-dual}
\end{equation}
In this range, $g_m\in\mathbb Z$ implies $\mu_m=\lfloor R_A n_m\rfloor+g_m-1$, and the equality $K_m=\mu_m+1-g_m=\lfloor R_A n_m\rfloor$ holds exactly.
Thus, $R_A$ is the limiting rate of the AG code $A_m$.

For each $m$, fix an $\FQ$-linear isomorphism $E_{A_m}:\FQ^{K_m}\to A_m$ for encoding.
Specifically, construct a basis of $L(\mu_mP_{\infty,m})$ using the method of
\cite{ShumEtAl2001}, and use its evaluation vectors at the points specified above as the rows of a generator matrix.
For fixed $Q$, this isomorphism $E_{A_m}$ can be computed from the defining equations of the tower in polynomial time.

\subsection{The outer subcode and the distance theorem}
\label{subsec:gs-construction}
\label{subsec:gs-distance}
Use the GS base code $A_m$ with $R_A$ given by \eqref{eq:gs-base-rate}.
Here $k_0=\Theta(\eps^{-1}\log(1/\eps))$, so
$\gamma_Q=o_\eps(R_A)$ and $\gamma_Q<R_A<1-\gamma_Q$ for sufficiently small $\eps$.
To obtain binary rate tending to $\eps^2$, set
\begin{equation}
k_m:=\left\lfloor\frac{\eps^2 n_m}{\rho_{\mathrm{in}}}\right\rfloor,
\qquad N_m:=n_mn_0.
\label{eq:outer-design-dimension}
\end{equation}
Since $\eps^2/\rho_{\mathrm{in}}=\Theta(\eps)$ and
$R_A/(\eps^2/\rho_{\mathrm{in}})\to\infty$ as $\eps\to0$, we have
$1\leq k_m\leq K_m$ for each sufficiently small $\eps$ and all sufficiently
large $m$.
The binary rate satisfies the exact bound
\begin{equation}
R_m:=\rho_{\mathrm{in}}\frac{k_m}{n_m},
\qquad
0\leq\eps^2-R_m<\frac{\rho_{\mathrm{in}}}{n_m}
=\frac{k_0}{N_m}.
\label{eq:gs-rate}
\end{equation}
In particular, $R_m\to\eps^2$ as $m\to\infty$ for fixed $\eps$.
Let $r_m\geq2$ be an even integer nearest to $\lambda_{\beta_\eps}N_m$.

Choose a random $k_m$-dimensional subspace $U_m\leq\FQ^{K_m}$
from any distribution satisfying \eqref{H}, and independently choose
$\alpha_m$ uniformly from $(\FQ^\times)^{n_m}$.
Define
\begin{equation}
\begin{aligned}
A_{m,\alpha_m}&:=\diag(\alpha_m)A_m,\\
\Coutm(\alpha_m,U_m)&:=\diag(\alpha_m)E_{A_m}(U_m)
\subseteq A_{m,\alpha_m},\\
C_m(\alpha_m,U_m)&:=\Coutm(\alpha_m,U_m)\circ_\phi\Cin.
\end{aligned}
\label{eq:gs-subcode-construction}
\end{equation}
Thus $A_m$ is the fixed base code, $A_{m,\alpha_m}$ is the scaled base code,
and $\Coutm(\alpha_m,U_m)$ is the selected outer code.
The distance analysis requires only \eqref{H}.
The constructions in \Cref{ex:graph-families} also give short descriptions
and efficient encoding, as discussed in \Cref{sec:algorithms}.

\begin{table}[tbp]
\centering
\caption{Parameters for the GS construction. The bar on $\bar\rho_{\mathrm{in}}$ denotes the target before rounding; the outer dimension uses the actual inner rate.}
\label{tab:gs-parameters}
\begin{tabularx}{\textwidth}{@{}lX@{}}
\toprule
Quantity & Choice\\
\midrule
Rate and distance targets & $R=\eps^2$, $\delta_\eps=h_2^{-1}(1-\eps^2-\eps^4)$, $\beta_\eps=1-2\delta_\eps$\\
Inner code & $\tau=\eps^2$, $n_0=\lceil c_0\eps^{-2}\log_2(1/\eps)\rceil$, $\bar\rho_{\mathrm{in}}=\log_2(1+\beta_\eps)+2\eps^2$\\
Dimension and field & Even $k_0$ with $|k_0-\bar\rho_{\mathrm{in}}n_0|\leq2$; $\rho_{\mathrm{in}}=k_0/n_0$, $Q=2^{k_0}$\\
Base code rate & $R_A=c_A\eps\log_2(1/\eps)$\\
Outer dimension & $k_m=\lfloor\eps^2n_m/\rho_{\mathrm{in}}\rfloor$\\
Moment order & Even $r_m$ nearest to $\lambda_{\beta_\eps}N_m$\\
\bottomrule
\end{tabularx}
\end{table}

The constants and remainder bounds are uniform over the admissible inner codes and encoding maps.
The limit in $m$ is taken with $\eps$ fixed.

\begin{lemma}
\label{lem:gs-exponents}
For the preceding parameter choices,
\begin{align}
\frac1{N_m}\ln\mathsf T_0
&\leq-\eps^4\ln2
+O\!\left(\frac{\eps^{c_0+2}}{\log(1/\eps)}\right)+o_m(1),
\label{eq:gs-T0-exponent}\\
\frac1{N_m}\log_2\mathsf T_A
&\leq\eps^2\log_2(1/\eps)
\left[2\ln2-c_A\sqrt{\frac2{\ln2}}+o_\eps(1)\right]
+O(\eps^2)+o_m(1).
\label{eq:gs-TA-exponent}
\end{align}
\end{lemma}
\begin{proof}
Write $e_\eps:=\rho_{\mathrm{in}}-\bar\rho_{\mathrm{in}}$, so $|e_\eps|\leq2/n_0$.
The inner parameters give
\[
(\rho_{\mathrm{in}}-\tau)\ln2-\ln(1+\beta_\eps)
=(\eps^2+e_\eps)\ln2,
\qquad
n_0(\eps^2+e_\eps)\ln2\geq c_0\ln(1/\eps)-2\ln2.
\]
Hence \eqref{eq:Gamma-at-saddle} implies
\begin{equation}
\Gamma_{\rho_{\mathrm{in}}-\tau,n_0}(s_{\beta_\eps})-1\leq4\eps^{c_0},
\qquad
\frac1{n_0}\ln\Gamma_{\rho_{\mathrm{in}}-\tau,n_0}(s_{\beta_\eps})
=O\!\left(\frac{\eps^{c_0+2}}{\log(1/\eps)}\right).
\label{eq:gs-gamma-bound}
\end{equation}
Since $R_m\leq\eps^2$ and $\mathsf I(\beta_\eps)=(\eps^2+\eps^4)\ln2$,
substitution into \eqref{eq:T0-exponent-general} proves
\eqref{eq:gs-T0-exponent}.
The error from rounding $r_m$ and from Stirling's formula tends to zero as $m\to\infty$.

For the second term, \eqref{eq:TA-exponent-general} and \eqref{eq:gs-dim-dual} give
\[
\frac1{N_m}\log_2\mathsf T_A
\leq R_m+\frac{r_m}{N_m}\log_2\frac1{\beta_\eps}
-\rho_{\mathrm{in}}\bigl(R_A-\gamma_Q+o_m(1)\bigr)\log_Q(Q-1).
\]
We have
\begin{align*}
\beta_\eps&=\sqrt{2\ln2}\,\eps+O(\eps^3),\\
\rho_{\mathrm{in}}&=\sqrt{\frac2{\ln2}}\,\eps+O(\eps^2),\\
\frac{r_m}{N_m}&=2\ln2\,\eps^2+O(\eps^4)+o_m(1).
\end{align*}
Also, $\gamma_Q=o_\eps(R_A)$ and $\log_Q(Q-1)=1-o_\eps(1)$.
Substitute these estimates and $R_A=c_A\eps\log_2(1/\eps)$ to obtain
\eqref{eq:gs-TA-exponent}.
The bounds use the inner code only through its parameters and the $\tau$-nice condition.
\end{proof}

For the following theorem, use the parameters and construction in
\Cref{subsec:gs-parameters,subsec:gs-construction}, with $c_0,c_A$
satisfying \eqref{eq:gs-ca-threshold}.
For each sufficiently small $\eps>0$, fix any $\tau$-nice inner code
and any associated encoding isomorphism.
The probability is over $(\alpha_m,U_m)$.

\begin{theorem}[Distance guarantee for the GS construction]
\label{thm:gs-gv}
\label{thm:gs-low-rate-gv}
There are constants $\eps_0,c>0$, independent of $\eps$, $m$,
and the chosen inner code and encoding map, such that the following holds.
For every $0<\eps<\eps_0$ and all sufficiently large $m$, depending on $\eps$,
the binary rate satisfies \eqref{eq:gs-rate}, and
\begin{equation}
\Prob_{\alpha_m,U_m}[C_m\ \mathrm{is}\ \beta_\eps\text{-}\mathrm{balanced}]
\geq1-2^{-c\eps^4N_m}.
\label{eq:gs-distance-probability}
\end{equation}
\end{theorem}
\begin{proof}
The choice of $c_A$ makes the leading coefficient in
\eqref{eq:gs-TA-exponent} strictly negative.
Thus, for some $c_1>0$ and sufficiently small $\eps$ and large $m$,
\[
\mathsf T_A\leq2^{-c_1\eps^2\log_2(1/\eps)N_m}.
\]
Since $c_0>2$, the error in \eqref{eq:gs-T0-exponent} is $o_\eps(\eps^4)$.
Taking $m$ sufficiently large after fixing $\eps$ gives
$\mathsf T_0\leq2^{-c_2\eps^4N_m}$ for some $c_2>0$.
Apply \eqref{eq:intro-balanced-bound} and decrease the exponent constant to absorb the sum of the two terms.
Thus $C_m$ is $\beta_\eps$-balanced with the stated probability.
\end{proof}
For the explicit distance bound, use
\[
1-h_2((1-\beta)/2)\geq\frac{\beta^2}{2\ln2}
\qquad(0\leq\beta<1).
\]
The function on the left has value and derivative zero at $\beta=0$,
and its second derivative is $1/((1-\beta^2)\ln2)\geq1/\ln2$.
Together with \eqref{eq:gs-target-distance}, this gives
\[
\frac{\beta_\eps}{2}
\leq\sqrt{\frac{\ln2}{2}}\,\eps\sqrt{1+\eps^2}
\leq\sqrt{\frac{\ln2}{2}}\left(\eps+\frac{\eps^3}{2}\right).
\]
Hence, the distance target also gives the explicit lower bound
\begin{equation}
\delta(C_m)\geq\delta_\eps
\geq\frac12-\sqrt{\frac{\ln2}{2}}
\left(\eps+\frac{\eps^3}{2}\right)
\label{eq:gs-first-order-gv}
\end{equation}
on the event in \eqref{eq:gs-distance-probability}.

If the inner code is also sampled independently, the total failure probability is at most
$n_02^{-\tau n_0}+2^{-c\eps^4N_m}$.
The first term tends to zero as $\eps\to0$; the exponential estimate in $N_m$ applies after fixing a $\tau$-nice inner code.
Equation \eqref{eq:gs-rate} and the distance bound prove \Cref{thm:intro-ag}.

\subsection{A simultaneous guarantee for the inner encoding maps}
\label{subsec:simultaneous-inner}

The constant in \Cref{thm:gs-gv} is independent of the chosen inner code and encoding map.
Since $n_0$ and $k_0$ remain fixed as $m\to\infty$, a finite union bound gives an outer code that works for all maps with $\tau$-nice images.

Under the assumptions of \Cref{thm:gs-gv}, let $\mathfrak F_\eps$ be the set of injective $\Ftwo$-linear maps $\varphi:\FQ\to\Ftwo^{n_0}$ with $\tau$-nice images.
Write
\[
C_m^\varphi(\alpha_m,U_m):=
\Coutm(\alpha_m,U_m)\circ_\varphi\varphi(\FQ).
\]

\begin{corollary}
\label{cor:fixed-outer-random-inner}
For all sufficiently large $m$,
\begin{equation}
\Prob_{\alpha_m,U_m}\!\left[\exists\varphi\in\mathfrak F_\eps:
C_m^\varphi(\alpha_m,U_m)\ \mathrm{is\ not}\ \beta_\eps\text{-}\mathrm{balanced}\right]
\leq2^{n_0k_0-c\eps^4N_m}.
\label{eq:simultaneous-seed}
\end{equation}
Consequently, for each sufficiently large $m$ one can fix a seed
$(\alpha_m^*,U_m^*)$, depending on $\eps$ and $m$, for which
$C_m^\varphi(\alpha_m^*,U_m^*)$ is $\beta_\eps$-balanced for every
$\varphi\in\mathfrak F_\eps$. For this fixed outer code, a uniformly random
injective $\Ftwo$-linear map $\varphi:\FQ\to\Ftwo^{n_0}$ satisfies
\[
\Prob_\varphi\!\left[
\delta\bigl(C_m^\varphi(\alpha_m^*,U_m^*)\bigr)<\delta_\eps
\right]\leq n_0 2^{-\tau n_0}.
\]
This upper bound tends to zero as $\eps\to0$.
\end{corollary}
\begin{proof}
There are at most $2^{n_0k_0}$ binary linear maps from $\FQ$ to $\Ftwo^{n_0}$.
Apply \eqref{eq:gs-distance-probability} to each $\varphi\in\mathfrak F_\eps$ and take a union bound.
For fixed $\eps$, the resulting upper bound tends to zero as $m\to\infty$, proving the existence assertion.
The image of a uniformly random injective map is a uniformly random binary $[n_0,k_0]$ code.
Its probability of failing to be $\tau$-nice is at most $n_0 2^{-\tau n_0}$ by \eqref{eq:tau-sampling-failure}.
Every map whose image is $\tau$-nice succeeds for the chosen outer code, which proves the last assertion.
\end{proof}

\section{Algorithmic properties of the construction}
\label{sec:algorithms}

We estimate the randomness required by the construction in \Cref{sec:ag-construction} and describe its encoding and decoding algorithms.
The distance theorem allows any distribution satisfying \eqref{H}; the complexity statements below use the constructions in \Cref{ex:graph-families}, or more generally subspaces with bases that can be computed in polynomial time.

\subsection{Randomness}
\label{subsec:short-seed}
In the construction using Toeplitz matrices, the seed consists of the matrix defining $U_Z$ and the coordinate scalings $\alpha\in(\FQ^\times)^n$.
Let $S_{\mathrm{seed}}$ denote its Shannon entropy in bits.
If $s_0:=K-k>0$, the matrix is determined by $K-1$ independent uniform elements of $\FQ$; if $s_0=0$, the subspace is deterministic.
Thus,
\begin{equation}
\frac{S_{\mathrm{seed}}}{N}
=
\rho_{\mathrm{in}}
\left(
\log_Q(Q-1)
+
\1[s_0>0]\frac{K-1}{n}
\right).
\label{eq:seed-entropy}
\end{equation}
Representing each field element by $k_0$ bits gives a seed description of length
\[
B_{\mathrm{desc}}
\leq
\bigl(n+\1[s_0>0](K-1)\bigr)k_0
\leq2\rho_{\mathrm{in}}N.
\]
These quantities exclude the descriptions of the fixed base code $A$, its encoding isomorphism $E_A$, and the inner encoding map $\phi$.

For the parameters of \Cref{thm:gs-gv}, we have $\rho_{\mathrm{in}}=\Theta(\eps)$ and $K/n=O(\eps\log_2(1/\eps))$.
Hence both the entropy and the description length are $O(\eps N)$, with the choice of $U_Z$ contributing $O(\eps^2\log_2(1/\eps))$ bits per binary coordinate.
The construction using multiplication in an extension field satisfies the same bounds.
An exactly uniform seed can be sampled using $O(\eps N)$ unbiased random bits in expectation by rejecting zero when choosing each $\alpha_i$.

\subsection{Encoding}
\label{sec:seed-and-encoding}
\label{subsec:encoding}

Let $G_A\in\FQ^{K\times n}$ represent $E_A$ with messages written as row vectors.
The outer encoder in the Toeplitz construction is
\[
u\longmapsto u[I_k\mid T^\top]G_A\diag(\alpha),
\qquad u\in\FQ^k.
\]
For the construction using multiplication in an extension field, replace $T^\top$ by a matrix representing $H_z$.
Applying $\phi:\FQ\to\Cin$ to each outer coordinate gives the binary encoder.
Thus, if $A$ can be encoded in polynomial time, so can the concatenated code.
For the GS codes, the required encoding maps are computed as described in \Cref{sec:garcia-stichtenoth}.

\subsection{Decoding inherited from the base code}
\label{subsec:inherited-decoding}

Suppose that $A$ has a decoder that corrects $e$ errors and $s$ erasures whenever $2e+s<D_A^{\mathrm{dec}}$, where $0<D_A^{\mathrm{dec}}\leq d(A)$.
To decode $\Cout(\alpha,Z)\subseteq A_\alpha$, multiply each unerased received coordinate by $\alpha_i^{-1}$ and apply the decoder for $A$.
If the result $a$ belongs to $E_A(U_Z)$, return $\diag(\alpha)a$; otherwise, declare failure.
Nonzero scaling preserves the error and erasure locations, so this decoder has the same correction guarantee.
The required generator matrix and membership tests are obtained by linear algebra from the seed and $G_A$.

Let $d_{\mathrm{in}}:=d(\Cin)$.
Forney's generalized minimum distance (GMD) decoding therefore corrects any $t$ binary errors in $\Cout(\alpha,Z)\circ_\phi\Cin$ whenever
\begin{equation}
t<\frac{d_{\mathrm{in}}D_A^{\mathrm{dec}}}{2},
\label{eq:inherited-decoding-radius}
\end{equation}
for every $\alpha,Z$ \cite{Forney1966GMD,GuruswamiIndyk2004}.
GMD decoding finds a nearest codeword in each inner block, orders the blocks by reliability, and tries at most $n+1$ erasure patterns, retaining only those with $s<D_A^{\mathrm{dec}}$.
Each candidate is reencoded and checked against the received binary word within the radius in \eqref{eq:inherited-decoding-radius}.
There is at most one such candidate, since the concatenated code has minimum distance at least $d_{\mathrm{in}}D_A^{\mathrm{dec}}$.
Given the generator matrix, membership testing, reencoding, and distance verification take polynomial time.
For a fixed inner code and encoding map, exhaustive decoding of one inner block takes $O(Qn_0)$ time, which is independent of $N$.
Thus the concatenated decoder runs in polynomial time if the decoder for $A$, including the preprocessing for each erasure pattern, does so.

For $A=C_L(D,G)$, suppose that the curve has an $\FQ$-rational point $P$ outside the support of $D$.
With auxiliary point $P$, the decoder of Lee, Bras-Amor{\'o}s, and O'Sullivan corrects $e$ errors whenever $2e<n-\deg G$ \cite[Propositions~13 and~15]{LeeBrasAmorosOSullivan2014}.
For $s$ erasures, apply this algorithm to $C_L(D',G)$, where $D'$ is the sum of the $n-s$ remaining evaluation points.
If $2e+s<n-\deg G$, then $n-s>\deg G$, so evaluation on these points is injective.
The recovered function can therefore be evaluated at all original points to recover the transmitted word.
Hence we may take $D_A^{\mathrm{dec}}=n-\deg G$.

For the GS codes, the required preprocessing also takes polynomial time.
Put $P=P_{\infty,m}$, $g=g_m$, and $n'=\deg D'$.
Choose a function $x$ whose only pole is at $P$, of least positive order $h\leq g+1$.
The standard bases over $\FQ[x]$ for $\mathcal R=\bigcup_{b\geq0}L(bP)$ and its ideal of functions vanishing on $D'$ are obtained by choosing functions of least pole order in each residue class modulo $h$ \cite[Section~II]{LeeBrasAmorosOSullivan2014}.
Riemann--Roch bounds these pole orders by $2g+h-1$ and $n'+2g+h-1$, respectively.
Interpolation functions are obtained from the surjective evaluation map $L((n'+2g)P)\to\FQ^{n'}$, and the vanishing conditions are imposed by taking kernels of evaluation maps.
Since $g=O(n_m)$, the basis functions and their products lie in Riemann--Roch spaces whose divisor degrees are $O(n_m)$.
The explicit function representations and basis computations on the tower \cite{ShumEtAl2001}, followed by linear algebra, give a polynomial bound for fixed $Q$, uniformly over the erasure patterns.
Together with the bounds on the remaining operations in \cite[Section~III.C]{LeeBrasAmorosOSullivan2014}, this includes the preprocessing in the running time of each decoding trial.

For the construction in \Cref{thm:gs-gv}, with either distribution from \Cref{ex:graph-families}, we obtain the following result.

\begin{corollary}[A decoding guarantee for finite GS codes]
\label{cor:gs-efficient-decoding}
For every sufficiently small fixed $\eps>0$, there exists a $\tau$-nice inner code $\Cin$ satisfying
\[
\frac{d(\Cin)}{n_0}\geq\frac12-\sqrt{\rho_{\mathrm{in}}}.
\]
For any such code and any isomorphism $\phi:\FQ\to\Cin$, for all sufficiently large $m$ and every outer seed $(\alpha_m,Z_m)$, the concatenated code can uniquely correct any $t$ binary errors satisfying
\begin{equation}
\frac{t}{N_m}<\vartheta_{\eps,m}
:=\frac12\left(\frac12-\sqrt{\rho_{\mathrm{in}}}\right)
\left(1-\frac{\mu_m}{n_m}\right)
\label{eq:gs-decoding-radius}
\end{equation}
in time polynomial in $N_m$.
\end{corollary}

\begin{proof}
For a uniformly random binary $[n_0,k_0]$ linear code, the probability of violating \eqref{eq:nice} is at most $n_02^{-\tau n_0}$.
For sufficiently small $\eps$, the entropy inequality used in the proof of \Cref{thm:gs-low-rate-gv} gives
$1-h_2(1/2-\sqrt{\rho_{\mathrm{in}}})\geq2\rho_{\mathrm{in}}/\ln2$.
A union bound over nonzero words of weight below $n_0(1/2-\sqrt{\rho_{\mathrm{in}}})$ therefore gives
\[
\Prob\!\left[d(\Cin)/n_0<1/2-\sqrt{\rho_{\mathrm{in}}}\right]
\leq2^{-(2/\ln2-1)k_0}.
\]
The sum of these two failure bounds is $o_\eps(1)$, so an inner code satisfying both conditions exists for sufficiently small $\eps$.

For $A_m$, we may take $D_{A_m}^{\mathrm{dec}}=n_m-\mu_m$.
Thus \eqref{eq:inherited-decoding-radius} gives
\[
\frac{d(\Cin)D_{A_m}^{\mathrm{dec}}}{2N_m}
=\frac12\frac{d(\Cin)}{n_0}\left(1-\frac{\mu_m}{n_m}\right)
\geq\vartheta_{\eps,m}.
\]
The polynomial running time follows from the preceding GMD and preprocessing bounds, since fixed $\eps$ allows us to fix $\Cin$ and $\phi$ independently of $m$.
\end{proof}

Since $\mu_m/n_m\to R_A+\gamma_Q$ for fixed $\eps$, the parameter estimates in \Cref{sec:ag-construction} give
\[
\lim_{m\to\infty}\vartheta_{\eps,m}
=\frac12\left(\frac12-\sqrt{\rho_{\mathrm{in}}}\right)(1-R_A-\gamma_Q)
=\frac14-O(\sqrt{\eps})
\qquad(\eps\to0).
\]
On the event in \eqref{eq:gs-distance-probability}, the lower bound on $\delta(C_m)/2$ has the form $1/4-O(\eps)$.
The larger deficit in the guaranteed GMD radius comes from the term $\sqrt{\rho_{\mathrm{in}}}$ in \eqref{eq:gs-decoding-radius}.

\section{Discussion}
\label{sec:limitations}
\label{sec:conclusion}
We compare the construction with unrestricted random outer codes,
explain the allocation of the inner and outer rates, and discuss the remaining questions.

\subsection{Comparison with random linear outer codes}
\label{subsec:random-outer-application}
Taking $A=\FQ^n$ removes the restriction to a proper base code.
Since $A^\perp=\{0\}$, the second term vanishes.
For $U$ chosen uniformly from $\operatorname{Gr}_Q(k,n)$, condition~\eqref{H} holds, and each map $U\mapsto\diag(\alpha)E_A(U)$ preserves the uniform distribution.
Thus, for every $0<\beta<1$ and every even $r\geq2$, \eqref{eq:intro-balanced-bound} gives
\begin{equation}
\Prob_U\!\left[U\circ_\phi\Cin\  \mathrm{is\ not}\ \beta\text{-}\mathrm{balanced}\right]
\leq\mathsf T_0.
\label{eq:uniform-outer-transfer}
\end{equation}

To recover the distance guarantee of \cite[Theorem~3]{DoronMosheiffWootters2024},
take both constituent rates equal to $\eps$ and $\log_2 Q=\Omega(\eps^{-3})$.
Then $n_0=(\log_2 Q)/\eps$, so $\tau=n_0^{-1/2}=o(\eps)$ and
$n_0\eps^2\to\infty$ as $\eps\to0$. For a $\tau$-nice inner code take
$\beta=2\eps$, $s=\eps/2$, and $r$ to be the nearest even integer to $\eps^2N$.
With $a=1$, $b=2$, and $z=1/2<\ln2$, \eqref{eq:rate-allocation-exponent}
below gives $N^{-1}\ln\overline{\mathsf T}_0=\eps^2(\ln2-7/8+o(1))$.
The coefficient $\ln2-7/8$ is negative, and a uniformly random inner code
fails \eqref{eq:nice} with probability at most $n_0 2^{-\sqrt{n_0}}$.
The resulting binary code has rate $\eps^2$. By \eqref{eq:uniform-outer-transfer},
the probability over the two independent codes that its relative distance is
below $1/2-\eps$ is at most $n_0 2^{-\sqrt{n_0}}+e^{-c\eps^2N}$ for
some $c>0$ and all sufficiently small $\eps$.

For the inner parameters in \eqref{eq:inner-design}, fix a $\tau$-nice inner
code and an encoding isomorphism $\phi:\FQ\to\Cin$ as in
\Cref{subsec:gs-parameters}, and take
$k=\lfloor\eps^2n/\rho_{\mathrm{in}}\rfloor$.
Then the binary rate tends to $\eps^2$ as $n\to\infty$, and the estimate of
$\mathsf T_0$ in \Cref{lem:gs-exponents} gives the distance target
$h_2^{-1}(1-\eps^2-\eps^4)$ with failure probability at most
$2^{-c\eps^4nn_0}$ for sufficiently large $n$.

\subsection{Allocation of the inner and outer rates}
\label{subsec:rate-allocation}
We explain how the inner rate affects the explicit upper bound
$\overline{\mathsf T}_0$ used in \Cref{sec:zero-tuple-moments}.
Doron, Mosheiff, and Wootters take both constituent rates to be $\eps$
in their analysis of random linear outer codes
\cite[Theorem~3]{DoronMosheiffWootters2024}.
The earlier analysis of Barg, Justesen, and Thommesen treats the inner
rate and the total rate as separate parameters \cite[Section~I]{BargJustesenThommesen2001}.
Thus equal rates are one relevant choice, rather than a common assumption of previous work.

For the comparison below, consider a family of $\tau$-nice inner codes with
$0<\tau<\rho_{\mathrm{in}}<1$, $R=\eps^2$,
$\rho_{\mathrm{in}}=a\eps+o(\eps)$, and $\tau=o(\eps)$, where $a>0$ is fixed.
Assume that $n_0\eps^2\to\infty$.
Take $\beta=b\eps+o(\eps)$ with fixed $b>0$,
$s=z\eps+o(\eps)$ with fixed $0<z<a\ln2$, and even $r$ such that
$r/N=\beta s+o(\eps^2)$.
Let $b_{\mathrm{crit}}(a)$ be the infimum of the $b>0$ for which
$\ln2-bz+z^2/2<0$ for some $0<z<a\ln2$.
Put $a_*:=\sqrt{2/\ln2}$ and $b_{\mathrm{GV}}:=\sqrt{2\ln2}$
\begin{proposition}
\label{prop:rate-split-threshold}
Under these assumptions,
\begin{equation}
\frac1N\ln\overline{\mathsf T}_0
=\eps^2\left(\ln2-bz+\frac{z^2}{2}\right)+o(\eps^2).
\label{eq:rate-allocation-exponent}
\end{equation}
Moreover,
\[
b_{\mathrm{crit}}(a)=
\begin{cases}
\displaystyle\frac1a+\frac{a\ln2}{2},&a<\sqrt{2/\ln2},\\[1ex]
\sqrt{2\ln2},&a\geq\sqrt{2/\ln2}.
\end{cases}
\]
\end{proposition}

\begin{proof}
The zero word gives $n_0^{-1}\ln\mathsf M_{\Cin}(s)\geq s-\rho_{\mathrm{in}}\ln2$,
which motivates the restriction $z<a\ln2$.
In this range,
\[
(\rho_{\mathrm{in}}-\tau)\ln2-\ln(1+\tanh s)
=(a\ln2-z)\eps+o(\eps)>0.
\]
Equation \eqref{eq:Gamma-general} and $n_0\eps^2\to\infty$ imply
$n_0^{-1}\ln\Gamma=o(\eps^2)$.
Since $r\asymp N\eps^2\to\infty$, Stirling's formula gives
\[
\frac1N\ln\frac{r!}{(\beta sN)^r}
=-\beta s+o(\eps^2).
\]
Together with $\ln\cosh s=z^2\eps^2/2+o(\eps^2)$, this proves
\eqref{eq:rate-allocation-exponent}.

The supremum of $bz-z^2/2$ over $0<z<a\ln2$ is $b^2/2$ if
$b\leq a\ln2$, and $ab\ln2-a^2(\ln2)^2/2$ otherwise.
Equating this supremum to $\ln2$ gives $b_{\mathrm{crit}}(a)$.
If $a<a_*$, every admissible $z$ satisfies
\[
\ln2-b_{\mathrm{GV}}z+\frac{z^2}{2}
=\frac{(b_{\mathrm{GV}}-z)^2}{2}
\geq\frac{(b_{\mathrm{GV}}-a\ln2)^2}{2}>0.
\]
Thus the upper bound diverges when $\beta=(b_{\mathrm{GV}}+o(1))\eps$,
because $N\eps^2\to\infty$.

If $a>a_*$, take $s=(b_{\mathrm{GV}}+\nu_\eps)\eps$ and let $r$ be the
nearest even integer to $\beta sN$, where $\beta=(b_{\mathrm{GV}}+\nu_\eps)\eps$.
Choose $\nu_\eps\to0$ slowly enough that $N\eps^2\nu_\eps\to\infty$
and the remainders in the exponent are $o(\eps^2\nu_\eps)$.
The normalized exponent is then
$-\eps^2(b_{\mathrm{GV}}\nu_\eps+\nu_\eps^2/2)+o(\eps^2\nu_\eps)$,
so $\overline{\mathsf T}_0\to0$.
At $a=a_*$, the optimum is on the excluded boundary $z=a\ln2$,
which explains the final qualification.
\end{proof}
In particular, if $a<a_*$, the stated range of $s,r$ cannot make
$\overline{\mathsf T}_0\to0$ when $\beta=(b_{\mathrm{GV}}+o(1))\eps$.
If $a>a_*$, there is a sequence $\nu_\eps>0$ tending to zero and admissible
$s,r$ for which $\beta=(b_{\mathrm{GV}}+\nu_\eps)\eps$ and
$\overline{\mathsf T}_0\to0$.

For equal constituent rates, $a=1$, the threshold for the distance deficit
given by this estimate is
$b_{\mathrm{crit}}(1)/2=(1+\ln2/2)/2=0.673286\ldots$,
whereas the GV coefficient is $\sqrt{\ln2/2}=0.588705\ldots$.
This comparison concerns the bound, not the best distance attainable by such codes.
Our construction has
$\rho_{\mathrm{in}}\sim a_*\eps$ and
$\lim_m k_m/n_m\sim\eps/a_*$ as $\eps\to0$.
The estimates in \Cref{lem:gs-exponents} control the smaller terms needed
when the leading coefficient of the inner rate approaches $a_*$.

\subsection{Open problems}
\label{subsec:open-problems}
The construction gives the leading GV coefficient while retaining a prescribed
AG base code for encoding and decoding. 

First, the selection of the outer subcode and the coordinate scalings is random.
The simultaneous guarantee in \Cref{cor:fixed-outer-random-inner} proves the
existence of an outer code that works for every inner map with a $\tau$-nice
image, at each sufficiently large block length.
It remains to find such a choice deterministically, or to certify efficiently
that a given choice satisfies the distance bound.

Second, the alphabet is fixed as the block length grows, but increases as
the target rate tends to zero:
$Q=2^{\Theta(\eps^{-1}\log(1/\eps))}$.
Reducing this dependence while preserving the distance guarantee, the
$O(\eps N)$ description length, and the decoding guarantee would improve
the cost of exhaustive decoding of the inner code.

Third, GMD decoding guarantees correction of a fraction
$1/4-O(\sqrt\eps)$ of errors, whereas half the distance lower bound is
$1/4-O(\eps)$.
Methods using multiple candidates for the inner code and more information
than its minimum distance have been studied for concatenated codes
\cite{guruswami2002decoding}.
One may seek to combine them with list recovery for AG codes, which asks
for codewords agreeing with prescribed small lists on sufficiently many
coordinates, while requiring the outer word to belong to $E_A(U_Z)$
after undoing the scalings.
The present complexity guarantee is polynomial in the block length for
fixed $\eps$ and does not give a bound polynomial in $1/\eps$.

Finally, the estimate using the dual distance discards part of the weight
distribution of $A^\perp$.
For finite block lengths, the exact sum in \Cref{thm:moment-counts} or the
bounds in \hyperref[app:poisson-bounds]{Appendix A} may give a sharper
distance guarantee.

\appendix
\renewcommand{\thesection}{\Alph{section}}
\section{Additional bounds obtained by Poissonization}
\label{app:poisson-bounds}

This appendix develops another approach to bounding the additional term $\mathcal L_r(A,\Cin)$ arising from $A$.
Although it is not needed for our GS construction, the approach explains how our estimate relates to the Poissonization argument of Doron, Mosheiff, and Wootters and gives an additional bound through an evaluation of the weight enumerator.

For a fixed tuple length $r$, \Cref{thm:moment-counts} gives
\[
\mathcal L_r(A,\Cin)=\sum_{w=1}^n\pi_r(w)\theta_w(A).
\]
Following \cite[Section~1.2]{DoronMosheiffWootters2024}, we replace the fixed tuple length by a Poisson random variable (Poissonization) to obtain another representation in terms of weight enumerators.

For $t\geq0$, define
\[
    p_0(t)
    :=
    \frac1Q W_{\Cin}(1,e^{-2t}),
    \qquad
    \xi(t)
    :=
    \frac{Qp_0(t)-1}{Q-1},
\]
and
\[
    \Phi_{A,\Cin}(t)
    :=
    Q^{-K}W_A(1,\xi(t))-p_0(t)^n.
\]

\begin{lemma}
\label{lem:poisson-weight-enumerator}
Let $\mathsf J\sim\operatorname{Poi}(Nt)$.
Conditional on $\mathsf J=j$, choose $V$ uniformly from $\mathcal V_j$, and choose $\alpha\in(\FQ^\times)^n$ independently and uniformly.
Then
\begin{equation}
\Prob_{\mathsf J,V,\alpha}
\left[
    g_V\neq0,\;
    g_V\in A_\alpha^\perp
\right]
=
\Phi_{A,\Cin}(t).
\label{eq:poisson_weight_enumerator}
\end{equation}
Moreover, for every integer $r\geq1$,
\begin{equation}
    \mathcal L_r(A,\Cin)
    \leq
    3\sqrt r\,\Phi_{A,\Cin}(r/N).
\label{eq:depoisson}
\end{equation}
\end{lemma}

\begin{proof}
By the splitting property of the Poisson distribution, the counts at positions $(i,j)\in[n]\times[n_0]$ are independent $\operatorname{Poi}(t)$ random variables.
Their parities are therefore independent Bernoulli random variables with parameter $(1-e^{-2t})/2$.

For one outer coordinate, let
\[
    \mathsf G_t
    :=
    \sum_{j=1}^{n_0}\zeta_j\omega_j,
    \qquad
    \Prob[\zeta_j=1]
    =
    \frac{1-e^{-2t}}{2}.
\]
For $a\in\FQ$, set $\chi_a(y):=(-1)^{\Tr(ay)}$.
Independence of the variables $\zeta_j$ gives $\E[\chi_a(\mathsf G_t)]=e^{-2t\wt(\phi(a))}$.
Fourier inversion therefore yields
\[
\Prob[\mathsf G_t=0]
=\frac1Q\sum_{a\in\FQ}e^{-2t\wt(\phi(a))}
=\frac1QW_{\Cin}(1,e^{-2t})=p_0(t).
\]
Thus, $p_0(t)$ is a probability, and the term $a=0$ also gives $p_0(t)\geq Q^{-1}$.
Hence, $0\leq\xi(t)\leq1$.

The $n$ outer coordinates are independent.
After multiplication by independent uniform elements of $\FQ^\times$, each coordinate equals $0$ with probability $p_0(t)$ and takes each nonzero value with probability $(1-p_0(t))/(Q-1)$.
Moreover, $g_V\in A_\alpha^\perp$ is equivalent to $\diag(\alpha)g_V\in A^\perp$.
Summing over $A^\perp\setminus\{0\}$ therefore gives
\[
\Prob[
    g_V\neq0,\;
    g_V\in A_\alpha^\perp
]
=
W_{A^\perp}
\left(
    p_0(t),
    \frac{1-p_0(t)}{Q-1}
\right)
-
p_0(t)^n.
\]
Since $p_0(t)+(Q-1)(1-p_0(t))/(Q-1)=1$ and $p_0(t)-(1-p_0(t))/(Q-1)=\xi(t)$, the MacWilliams identity \cite[Chapter~5]{MacWilliamsSloane1977} yields \eqref{eq:poisson_weight_enumerator}.

On the other hand, conditioning on $\mathsf J=j$ gives
\[
    \Phi_{A,\Cin}(t)
    =
    e^{-Nt}
    \sum_{j=0}^\infty
    \frac{(Nt)^j}{j!}
    \mathcal L_j(A,\Cin).
\]
Setting $t=r/N$ and retaining only the term $j=r$, we obtain
\[
    \Phi_{A,\Cin}(r/N)
    \geq
    \Prob[\operatorname{Poi}(r)=r]
    \mathcal L_r(A,\Cin).
\]
Stirling's formula gives
\[
    \Prob[\operatorname{Poi}(r)=r]^{-1}
    \leq
    \sqrt{2\pi r}\,e^{1/(12r)}
    <
    3\sqrt r,
\]
which proves \eqref{eq:depoisson}.
\end{proof}

Combining \Cref{thm:moment-counts,lem:poisson-weight-enumerator} gives, for every even $r\geq2$,
\[
\mathcal L_r(A,\Cin)
\leq\min\{\Theta_A(Q),\,3\sqrt r\,\Phi_{A,\Cin}(r/N)\}.
\]

The quantity $\Theta_A(Q)$ is the maximum, over $w$, of the fraction $\theta_w(A)$ of all vectors of weight $w$ that belong to the dual code, and is determined by the ordinary dual weight enumerator.
Poissonization, on the other hand, makes the contributions from different coordinates independent and evaluates the weight enumerator at a specified point.
The two approaches differ in how they weight and bound the contributions, rather than in the type of information they use about the outer code.

\bibliographystyle{plain}
\bibliography{main}

@article{Gilbert1952,
  author={Edgar N. Gilbert}, title={A Comparison of Signalling Alphabets},
  journal={Bell System Technical Journal}, volume={31}, number={3}, pages={504--522}, year={1952}}

@article{Varshamov1957,
  author={R. R. Varshamov}, title={Estimate of the Number of Signals in Error Correcting Codes},
  journal={Doklady Akademii Nauk SSSR}, volume={117}, pages={739--741}, year={1957}}

@article{GarciaStichtenoth1996,
  author={Arnaldo Garcia and Henning Stichtenoth},
  title={On the Asymptotic Behaviour of Some Towers of Function Fields over Finite Fields},
  journal={Journal of Number Theory}, volume={61}, number={2}, pages={248--273}, year={1996}}

@article{ShumEtAl2001,
  author={Kenneth W. Shum and Ilia Aleshnikov and P. Vijay Kumar and Henning Stichtenoth and Vinay Deolalikar},
  title={A Low-Complexity Algorithm for the Construction of Algebraic-Geometric Codes Better Than the {Gilbert--Varshamov} Bound},
  journal={IEEE Transactions on Information Theory}, volume={47}, number={6}, pages={2225--2241}, year={2001}}

@article{BargJustesenThommesen2001,
  author={Alexander Barg and J{\o}rn Justesen and Christian Thommesen}, title={Concatenated Codes with Fixed Inner Code and Random Outer Code},
  journal={IEEE Transactions on Information Theory}, volume={47}, number={1}, pages={361--365}, year={2001}}

@inproceedings{GuruswamiIndyk2004,
  author={Venkatesan Guruswami and Piotr Indyk}, title={Efficiently Decodable Codes Meeting {Gilbert--Varshamov} Bound for Low Rates},
  booktitle={Proceedings of the 15th Annual ACM-SIAM Symposium on Discrete Algorithms (SODA)}, pages={756--757}, year={2004}, note={Longer version appeared in Allerton 2003}}

@inproceedings{DoronMosheiffWootters2024,
  author={Dean Doron and Jonathan Mosheiff and Mary Wootters}, title={When Do Low-Rate Concatenated Codes Approach The {Gilbert--Varshamov} Bound?},
  booktitle={APPROX/RANDOM 2024}, series={LIPIcs}, volume={317}, pages={53:1--53:12}, year={2024}}

@InProceedings{BrehmIshaiResch2026,
  author =	{Brehm, Martijn and Ishai, Yuval and Resch, Nicolas},
  title =	{{Fast Bounded-Independence Functions and Their Duals}},
  booktitle =	{7th Conference on Information-Theoretic Cryptography (ITC 2026)},
  pages =	{8:1--8:23},
  series =	{Leibniz International Proceedings in Informatics (LIPIcs)},
  ISBN =	{978-3-95977-426-0},
  ISSN =	{1868-8969},
  year =	{2026},
  volume =	{385},
  editor =	{Dodis, Yevgeniy},
  publisher =	{Schloss Dagstuhl -- Leibniz-Zentrum f{\"u}r Informatik}
}

@article{BenAroyaTaShma2013,
  author={Avraham Ben-Aroya and Amnon Ta-Shma},
  title={Constructing Small-Bias Sets from Algebraic-Geometric Codes},
  journal={Theory of Computing}, volume={9}, number={5}, pages={253--272},
  year={2013}, doi={10.4086/toc.2013.v009a005}}

@article{UmedaSakakibaraKasahara1996,
  author={Takeshi Umeda and Katsumi Sakakibara and Masao Kasahara},
  title={Notes on the Average Binary Weight Enumerator of Generalized Algebraic-Geometric Codes},
  journal={IEICE Transactions on Fundamentals of Electronics, Communications and Computer Sciences},
  volume={E79-A}, number={9}, pages={1444--1446}, year={1996}}

@inproceedings{BrakensiekDharGopiZhang2025,
  title={{AG} codes achieve list decoding capacity over constant-sized fields},
  author={Brakensiek, Joshua and Dhar, Manik and Gopi, Sivakanth and Zhang, Zihan},
  booktitle={Proceedings of the 56th Annual ACM Symposium on Theory of Computing},
  pages={740--751},
  year={2024}
}

@inproceedings{CohenDoronGoldgraberManket2026,
  author={Gil Cohen and Dean Doron and Noam Goldgraber and Tomer Manket}, title={Tracing {AG} Codes: Toward Meeting the {Gilbert--Varshamov} Bound},
  booktitle={ICALP 2026}, series={LIPIcs}, volume={374}, pages={68:1--68:23}, year={2026}, eprint={2511.08788}, archivePrefix={arXiv}}

@article{LeeBrasAmorosOSullivan2014,
  author={Kwankyu Lee and Maria Bras-Amor{\'o}s and Michael E. O'Sullivan}, title={Unique Decoding of General {AG} Codes},
  journal={IEEE Transactions on Information Theory}, volume={60}, number={4}, pages={2038--2053}, year={2014}, eprint={1210.3101}, archivePrefix={arXiv}}

@book{Stichtenoth2009,
  author={Henning Stichtenoth},
  title={Algebraic Function Fields and Codes},
  edition={2nd},
  publisher={Springer},
  year={2009}}

@book{MacWilliamsSloane1977,
  author={Florence Jessie MacWilliams and Neil James Alexander Sloane},
  title={The Theory of Error-Correcting Codes},
  publisher={North-Holland},
  year={1977}}

@book{Forney1966,
  author    = {G. David Forney, Jr.},
  title     = {Concatenated Codes},
  series    = {M.I.T. Research Monograph},
  number    = {37},
  publisher = {The MIT Press},
  address   = {Cambridge, MA},
  year      = {1966}
}

@article{Zyablov1971,
  author  = {Victor V. Zyablov},
  title   = {An Estimate of the Complexity of Constructing Binary Linear Cascade Codes},
  journal = {Problems of Information Transmission},
  volume  = {7},
  number  = {1},
  pages   = {3--10},
  year    = {1971}
}

@article{Thommesen1983,
  author  = {Christian Thommesen},
  title   = {The Existence of Binary Linear Concatenated Codes with {Reed--Solomon} Outer Codes Which Asymptotically Meet the {Gilbert--Varshamov} Bound},
  journal = {IEEE Transactions on Information Theory},
  volume  = {29},
  number  = {6},
  pages   = {850--853},
  year    = {1983},
  doi     = {10.1109/TIT.1983.1056765}
}

@article{Forney1966GMD,
  author  = {G. David Forney, Jr.},
  title   = {Generalized Minimum Distance Decoding},
  journal = {IEEE Transactions on Information Theory},
  volume  = {12},
  number  = {2},
  pages   = {125--131},
  year    = {1966},
  doi     = {10.1109/TIT.1966.1053873}
}

@phdthesis{guruswami2002decoding,
  author       = {Venkatesan Guruswami},
  title        = {List Decoding of Error-Correcting Codes},
  school       = {Massachusetts Institute of Technology},
  year         = {2002},
  note         = {Winning Thesis of the 2002 ACM Doctoral Dissertation Competition},
  publisher    = {Springer},
  series       = {Lecture Notes in Computer Science},
  volume       = {3282}
}

\end{document}